\documentclass[a4paper,11pt,fleqn,usenatbib,useAMS]{article}
\usepackage{jcappub} 
\usepackage[sort&compress,numbers]{natbib}
\usepackage{graphicx}	
\usepackage{amsmath, amstext}	
\usepackage{amssymb}	
\usepackage{multicol}        
\usepackage{bm}		
\usepackage{cleveref}
\usepackage{pdflscape}	
\usepackage{xcolor}
\usepackage[figure,figure*]{hypcap}
\usepackage{float}
\usepackage{eso-pic}
\usepackage{tablefootnote}
\usepackage{aas_macros}
\usepackage[caption = false]{subfig}
\usepackage{hanging}
\usepackage{orcidlink}

\newcommand{\AMTL}{AMTL }
\newcommand{\AMTLnospace}{AMTL}
\newcommand{\mpch}{Mpc $h^{-1}$ }
\newcommand{\gpch}{Gpc $h^{-1}$ }

\usepackage[T1]{fontenc}
\usepackage{ae,aecompl}

\usepackage{newtxtext,newtxmath}

\title{Production of Alternate Realizations of DESI Fiber Assignment for Unbiased Clustering Measurement in Data and Simulations}

\collaboration{{DESI Collaboration}}
\author[1]{{J.~Lasker}\orcidlink{0000-0003-2999-4873},}
\author[2,3]{{A.~Carnero Rosell}\orcidlink{0000-0003-3044-5150},}
\author[4]{{A.~D.~Myers},}
\author[5,6,7]{{A.~J.~Ross}\orcidlink{0000-0002-7522-9083},}
\author[8]{{D.~Bianchi}\orcidlink{0000-0001-9712-0006},}
\author[9]{{M.~M.~S~Hanif}\orcidlink{0009-0006-2583-5006},}
\author[1]{{R.~Kehoe},}
\author[10]{{A.~de~Mattia},}
\author[4]{{L.~Napolitano}\orcidlink{0000-0002-5166-8671},}
\author[11,12,13]{{W.~J.~Percival}\orcidlink{0000-0002-0644-5727},}
\author[1]{{R.~Staten},}
\author[14]{{J.~Aguilar},}
\author[15]{{S.~Ahlen}\orcidlink{0000-0001-6098-7247},}
\author[16]{{L.~Bigwood},}
\author[17]{{D.~Brooks},}
\author[14]{{T.~Claybaugh},}
\author[16]{{S.~Cole}\orcidlink{0000-0002-5954-7903},}
\author[18]{{A.~de la Macorra}\orcidlink{0000-0002-1769-1640},}
\author[19]{{Z.~Ding}\orcidlink{0000-0002-3369-3718},}
\author[17]{{P.~Doel},}
\author[20,21]{{K.~Fanning}\orcidlink{0000-0003-2371-3356},}
\author[22,23]{{J.~E.~Forero-Romero}\orcidlink{0000-0002-2890-3725},}
\author[24,25,26]{{E.~Gaztañaga},}
\author[14]{{S.~Gontcho A Gontcho}\orcidlink{0000-0003-3142-233X},}
\author[27]{{G.~Gutierrez},}
\author[5,28,7]{{K.~Honscheid},}
\author[29]{{C.~Howlett}\orcidlink{0000-0002-1081-9410},}
\author[30]{{S.~Juneau},}
\author[14]{{A.~Kremin}\orcidlink{0000-0001-6356-7424},}
\author[14]{{M.~Landriau}\orcidlink{0000-0003-1838-8528},}
\author[31]{{L.~Le~Guillou}\orcidlink{0000-0001-7178-8868},}
\author[14]{{M.~E.~Levi}\orcidlink{0000-0003-1887-1018},}
\author[32,33]{{M.~Manera}\orcidlink{0000-0003-4962-8934},}
\author[30]{{A.~Meisner}\orcidlink{0000-0002-1125-7384},}
\author[34,33]{{R.~Miquel},}
\author[35]{{J.~Moustakas}\orcidlink{0000-0002-2733-4559},}
\author[36]{{E.~Mueller},}
\author[37]{{J.~Nie}\orcidlink{0000-0001-6590-8122},}
\author[38,39]{{G.~Niz}\orcidlink{0000-0002-1544-8946},}
\author[40]{{M.~Oh},}
\author[10,14]{{N.~Palanque-Delabrouille}\orcidlink{0000-0003-3188-784X},}
\author[14,41,42]{{C.~Poppett},}
\author[43]{{F.~Prada}\orcidlink{0000-0001-7145-8674},}
\author[44]{{M.~Rezaie}\orcidlink{0000-0001-5589-7116},}
\author[45]{{G.~Rossi},}
\author[46]{{E.~Sanchez}\orcidlink{0000-0002-9646-8198},}
\author[14]{{D.~Schlegel},}
\author[47,9]{{M.~Schubnell},}
\author[48]{{H.~Seo}\orcidlink{0000-0002-6588-3508},}
\author[30]{{D.~Sprayberry},}
\author[9]{{G.~Tarl\'{e}}\orcidlink{0000-0003-1704-0781},}
\author[18]{{M.~Vargas-Maga\~na}\orcidlink{0000-0003-3841-1836},}
\author[30]{{B.~A.~Weaver},}
\author[16]{{Michael~J.~Wilson},}
\author[19]{{Y.~Zheng},}

\emailAdd{spokespersons@desi.lbl.gov}
\affiliation{Affiliations are in Appendix \ref{sec:affiliations}}
\emailAdd{james.lasker3@gmail.com}
\emailAdd{jlasker@smu.edu}

\date{Last updated YYYY MM DD; in original form YYYY MM DD}

\abstract{A critical requirement of spectroscopic large scale structure analyses is correcting for selection of which galaxies to observe from an isotropic target list. This selection is often limited by the hardware used to perform the survey which will impose angular constraints of simultaneously observable targets, requiring multiple passes to observe all of them. In SDSS this manifested solely as the collision of physical fibers and plugs placed in plates. In DESI, there is the additional constraint of the robotic positioner which controls each fiber being limited to a finite patrol radius. A number of approximate methods have previously been proposed to correct the galaxy clustering statistics for these effects, but these generally fail on small scales.  To accurately correct the clustering we need to upweight pairs of galaxies based on the inverse probability that those pairs would be observed (Bianchi \& Percival 2017). This paper details an implementation of that method to correct the Dark Energy Spectroscopic Instrument (DESI) survey for incompleteness. To calculate the required probabilities, we need a set of alternate realizations of DESI where we vary the relative priority of otherwise identical targets.  These realizations take the form of alternate Merged Target Ledgers (\AMTLnospace), the files that link DESI observations and targets. We present the method used to generate these alternate realizations and how they are tracked forward in time using the real observational record and hardware status, propagating the survey as though the alternate orderings had been adopted. We detail the first applications of this method to the DESI One-Percent Survey (SV3) and the DESI year 1 data. We include evaluations of the pipeline outputs, estimation of survey completeness from this and other methods, and validation of the method using mock galaxy catalogs. }

\begin{document}

\maketitle
\flushbottom

\label{firstpage}




\section{INTRODUCTION}
\label{sec:intro}
Large scale spectroscopic sky surveys have been a primary way of studying the universe over the past twenty five years \cite{SDSSOverview, BOSSOverview, eBOSSOverview}. They obtain redshifts of many millions of galaxies in order to measure their position in three dimensional space and extract information regarding the clustering of galaxies and the dark matter fields which they trace throughout cosmic time. These measurements not only provide insight into the composition of the universe and the physics which drive the evolution of its expansion, but also the formation of the galaxies themselves.

The Dark Energy Spectroscopic Instrument (DESI) survey, like the extended Baryon Oscillation Spectroscopic Survey (eBOSS) and the Sloan Digital Sky Survey (SDSS) before it, utilizes a multi object spectrograph (MOS) which enables multiplexing of spectroscopic observations. However, DESI is able to multiplex at a scale $\sim$5x larger than that of eBOSS, capturing $\gtrsim$4000 spectra of astronomical objects per pointing. This increase in efficiency is largely due to the fact that SDSS and eBOSS hand-placed the fibers which fed their spectrographs into plates, while the fibers which feed the DESI spectrographs are controlled robotically. These positioners place  all 5000 fibers on targets within the focal plane \cite{DESIInstrument}. DESI also utilizes a dynamic exposure time calculator \cite{SurveyOps,Expcalc.Kirkby.2024} which takes information on observing conditions including seeing, transparency, and sky brightness to estimate the true signal-to-noise accumulated during an exposure. The capability to optimize exposure times and to construct new "plates" as needed, combined with a fivefold increase in available fibers and a $\sim 2.5x$ increase in the mirror area, allowed DESI to obtain $\sim$10x as many extragalactic redshifts in its first year as eBOSS did in its entire 5 year survey. 

While robotically controlled fibers have provided many gains to the DESI survey, they have not eliminated the problem of fiber collisions. A fiber collision occurs when two targets are close enough on the sky that two fibers and their plugs (in SDSS and eBOSS) or their positioners (in DESI) would have collided were a survey to attempt to simultaneously observe both \cite{Bla03}. DESI also has a limitation similar to that of fiber collisions where only one fiber may be placed within a single fiber patrol radius, with a small amount of overlap towards the edges of the patrol radii of neighboring fibers. This enforces uniform areal observation density within a single field of view which can only begin to reflect the non-uniform density of targets after multiple passes. Since they have similar effects on the ability to observe close pairs, we will refer to both effects collectively as 'fiber collisions.`

Additionally, robotic fiber positioners can suffer from time-variable issues which were not present in the hand-plugged fiber systems used by SDSS, such as losses to positioner power and software-based failures which interrupt the positioning process. Fiber-collision issues combine with more traditional problems such as sky subtraction failures, atmospheric variability, and redshift determination errors, to create incompleteness effects which vary spatially and temporally across the DESI survey.

Fiber collisions affect the smallest scale clustering since they are limited to angular scales on the sky of approximately $<1.48$~arcminutes which projects to a distance of 0.99 \mpch at an intermediate redshift of z = 1.0. Real time hardware events like power failures to the devices which control groups of fiber positioners (CAN busses) can also cause missing observations at larger sets of separations. These smaller scales are critical to analyses of the galaxy-halo connection. These analyses, whether they use Halo Occupation Distribution (HOD) methods \citep{Gao23, Pearl23, Rocher23,Smith23,Yuan23a, Yuan23b} or SubHalo Abundance Matching (SHAM) methods \citep{Yu23, Gao24, Prada24}, rely on clustering measurements at small scales to determine the placement and quantity of satellite galaxies.

Prior to the usage of pairwise inverse probability weighting, (e)BOSS and SDSS used have used several different methods to correct for incompleteness. The BOSS DR9 BAO analysis \citep{Anderson12} used the method developed by Guo, Zehavi, and Zheng in 2012 \cite{GuoZehaviZheng12} to correct for fiber collisions. These weights are determined by dividing the galaxy sample into those which are not subject to any fiber collision effects and those which are subject to those effects. The relative fraction of galaxies within the collided sample which are observed is calculated. Then the pair counts which include galaxies from the collided sample are upweighted based on that fraction. This method does not account for the fact that unobserved pairs are not statistically equivalent to those which are observed. The Reid et al 2014 \cite{Reid14} analysis of the growth of structure using small scale clustering utilizes a combination of angular upweighting on small scales, and nearest neighbor upweighting, which accounts for unobserved targets by adding one to a "close pair" weight for an observed target for each unobserved target which is closer to it than any other observed member of its target class, at large scales, and a truncation of the measured multipoles at small separation perpendicular to the line of sight.   Other standard SDSS and eBOSS analyses \cite{Anderson14, Reid16, Ross20, Lyke20, Raichoor21} have used nearest neighbor upweighting. This method assumes that all fiber collisions occur between galaxies within clusters or are otherwise at similar redshifts. 


To address this issue of measuring small scale clustering, Bianchi and Percival 2017 (hereafter BP17; \cite{BianchiAndPercival}) developed a new method of correction for incompleteness by upweighting each pair of galaxies rather than individual galaxies or the total number of paircounts. They devise a method of Pairwise Inverse Probability (PIP) weights which maintains consistent numbers of both weighted pair counts and weighted individual galaxy counts on all scales as long as there is some area of the survey footprint which is observed multiple times. 

Bianchi and Percival note that, despite the nominal possibility of computing such pairwise weights analytically, for modern surveys like DESI which will observe $\mathcal{O}(10^7)$ galaxies, such calculations will be practically impossible. They do note that these weights can be effectively estimated by randomly repeating the galaxy selection process a sufficient number of times and estimating the probability of each pair based on the frequency at which each pair is selected. The remainder of this paper will show a new method which allows us to estimate those probabilities, show its effect on the clustering measurements, and validate the weights using mocks.

Mohammad et al 2018 \cite{Mohammad18} showed the first application of the BP17 PIP weighting to data using the VIPERS survey. Additionally, they compute angular upweighting as described in Percival and Bianchi 2017 (hereafter PB17; \cite{PercivalAndBianchi}). This application to a slit-based survey is not strictly the same as the DESI application, but showed the validity of the BP17 PIP + PB17 angular upweighting method when applied to mocks and data. Mohammad et al 2020 \cite{Mohammad20} showed the first application of the BP17 PIP weighting + PB17 angular upweighting method to MOS data using the eBOSS DR16 catalogs. They used a version of the eBOSS tiling software which was modified to enable easier parallelization of fiber assignment across 1860 realizations of the eBOSS survey in which they varied only the initial random number generation used by the software. They store the results of the realizations as bitweights as recommended in BP17 and then use them to compute pairwise inverse probability weights. The authors then use EZ Mocks to validate that their application of PIP weighting successfully recovers one-halo clustering down to a scale of 0.1 \mpch.

In this paper we present a new method of calculating PIP weights to correct the DESI SV3 and Y1 clustering measurements for incompleteness. This method utilizing Alternate MTLs (AMTLs) is a nearly-maximally realistic monte carlo estimation of the likelihood that any target or collection of targets would be observed over an infinite repetition of the DESI experiment. This method incorporates more realism than prior treatements in eBOSS by utilizing real observation results, hardware states, and sequential changes in priority as overlapping tiles are observed. This method will enable unbiased measurement of clustering of galaxies down to scales $\leq 0.1$ \mpch.

In section \ref{sec:survey} we describe the DESI survey strategy and how this affects the observation of different classes of targets, in Section \ref{sec:mocks} we briefly discuss the mock galaxy catalogs used to validate this method, in Section \ref{sec:altmtl} we describe both the regular and alternate MTLs and the procedure for how we create and propagate the alternate MTLs, in Section \ref{sec:TPCFCode} we describe the DESI clustering code, \textsc{Pycorr} and how it implements these weights, in Section \ref{sec:results} we show results from the data and mocks to validate the method and show the differences with other methods of incompleteness determination, and finally in Section \ref{sec:discussion} we discuss the results and their applicability to other analyses within DESI.

\section{Survey}
\label{sec:survey}
The Dark Energy Spectroscopic Instrument  (DESI: \cite{DESIWhitePaper, DESIExperimentOverview1, DESIExperimentOverview2}) is the Stage IV successor to large scale sky surveys such as the Sloan Digital Sky Survey (SDSS: \cite{SDSSOverview}) and the (extended) Baryon Oscillation Spectroscopic Survey ((e)BOSS: \cite{eBOSSOverview}). The DESI experiment is currently in its third year of a projected five year program and has already recorded redshifts for over 30 million galaxies out to redshifts over $z\sim 3.5$. The primary goal of this survey is to further constrain cosmological parameters such as the dark energy equation of state parameter ($w$) and the Hubble Constant ($H_0$). The survey plans to do this by performing the most precise measurement of the Baryon Acoustic Oscillation (BAO) feature and of Redshift Space Distortions (RSD) in its clustering measurements. However, the science goals of DESI are not limited to dark energy. Already-published analyses from DESI include explorations of the Galaxy Halo Connection, measurements of the local stellar population in the Milky Way, studies of Quasistellar Object (QSO) astrophysics, and many tests of the survey and pipeline performances. This paper is part of a set of analyses surrounding the first Data Release of DESI (DR1, \cite{DESI2024.I.DR1}) which includes two point clustering measurements and validation \cite{DESI2024.II.KP3}, measurements of the BAO feature in Galaxies, QSOs, and the Lyman-$\alpha$ forest \cite{DESI2024.III.KP4, DESI2024.IV.KP6}, full shape measurements of Galaxy and QSO clustering \cite{DESI2024.V.KP5}, as well as cosmological constraints measured from the above features including tests of primordial non-gaussianity \cite{DESI2024.VI.KP7A, DESI2024.VII.KP7B, DESI2024.VIII.KP7C}.

DESI selects the galaxies and stars that it observes from a parent population of Milky Way targets (MWS; \cite{MWSTargetSelection}), Bright, low redshift, Galaxies (BGS; \cite{BGSTargetSelection}), Luminous Red Galaxies (LRGs; \cite{LRGTargetSelection}), Emission Line Galaxies (ELGs; \cite{ELGTargetSelection}), and Quasistellar Objects (Quasars/QSOs; \cite{QSOTargetSelection}). Targets are obtained via prior imaging of the DESI footprint\cite{DESILegacySurvey}. These targets are divided into two concurrent programs with BGS and MWS targets being observed during bright times or in other sub-optimal conditions while LRG, ELG, and QSO targets are observed only in sufficiently good dark time \citep{SurveyOps}. 

\begin{figure}
    \centering
    \label{fig:SV3RosettePattern}
    \includegraphics[scale = 0.5]{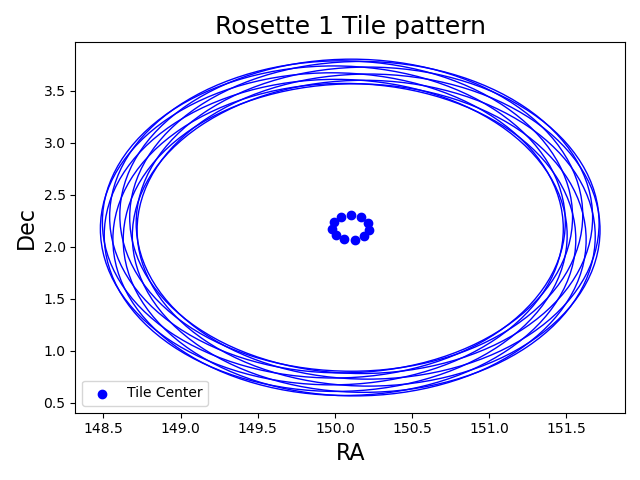}
    \caption{This plot shows the pattern in which an SV3 rosette is tiled, using Rosette 1 as an example. The points represent field centers and the circles represent the approximate (uncorrected for projection distortion) DESI FOV around each center.}
\end{figure}

\begin{figure}
    \centering
    \label{fig:SV3Flowchart}
    \includegraphics[]{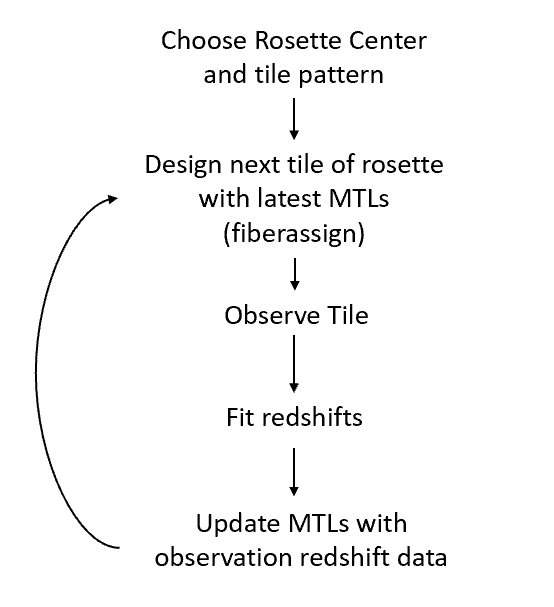}
    \caption{This flowchart sketches out the order in which SV3 operations happen. This highlights that MTLs are always updated in-between fiber assignment of one tile and fiber assignment of the next overlapping rosette tile.}
\end{figure}

During the One-Percent Survey (also referred to as SV3), these targets were observed within 20 rosettes \cite{SVOverview}. SV3 was designed to optimize the observation of complete samples of tracers over an area approximately equal to one percent of the final survey area. This survey also provided a final test for the survey operations procedures and software prior to the main survey. These rosettes were selected to be evenly distributed between the northern and southern galactic caps as well as to overlap with various fields of interest from CFHTLS, COSMOS, DES, GAMA, Euclid, ELAIS, GOODS, HSC, KiDS, VVDS, XBootes, and XDEEP as well as overlapping with the Coma cluster and the ecliptic pole. For each rosette, 12/10 (dark/bright) tiles were designed with tile centers spaced evenly around the rosette center at a distance of $0.12^{\circ}$ as shown in Figure~\ref{fig:SV3RosettePattern}. Observation of these rosettes followed a sequence where, for each rosette, one tile was designed and fiberassigned in the afternoon before observing or earlier, observed at night, and then the Merged Target Ledgers (MTLs\footnote{MTLs will be further explained in \S~\ref{sec:constructAMTL} and more details can be found in \S 6 of \cite{SurveyOps}.}) containing the targets which were observed in that tile\footnote{MTLs are divided into HEALPixels (https://healpix.jpl.nasa.gov/) with nside = 32 (pixel size $\sim 1.8^\circ$) for ease of reading and storage} were updated. Then the next overlapping tile on the same rosette was designed, incorporating the information from the previous night's observations, and observed in the same manner until all 12/10 tiles were observed. A flowchart of this process is shown in Figure~\ref{fig:SV3Flowchart}.

This strategy enabled high completeness of 96.2\% for MWS targets, 98.9\% for BGS targets, 98.6\% for LRG targets, 95.2\% for ELG targets, and 99.4\% for QSO targets within a range of rosette radii of 0.2 and 1.45 degrees\citep{SVOverview}. Despite the nearly 100\% completeness in that range, there is still an inner and outer region in which we can study the incompleteness measurements and weights from the \AMTL method. Additionally, some regions within the central complete area are still subject to fiber collision effects due to high target number density and nightly focal plane positioning issues. 

The observed SV3 spectra were processed by the DESI spectroscopic pipeline \cite{SpecPipe} in a homogeneous processing run denoted as `Fuji', with the resulting redshift and large-scale structure (LSS) catalogs used in this paper released in the early data release (EDR \cite{edr}). The results of the \AMTL pipeline we describe in this paper were included in the EDR LSS catalogs.

The main survey maintained the strategy of not observing overlapping areas of sky before updating the MTLs, but is no longer confined to rosettes. The tiles for the main survey were chosen based on the Hardin, Sloane, and Smith 2001 icosahedral tiling\footnote{http://neilsloane.com/icosahedral.codes/}\cite{Hardin01} and are often fiber assigned "on-the-fly" as survey observations are occurring. This allows for greater flexibility in where one can point the telescope during a night of observing due to changing conditions including wind and clouds. 
The main survey is ongoing, but the DESI data release 1 (DR1 \cite{DESI2024.I.DR1}) will include data from just over the first year of its operations, May 14, 2021 through June 14, 2022. Analyses of this dataset are ongoing and the resulting LSS catalogs have completeness 62\% in BGS, 69\% in LRG, 35\% in ELG, and 88\% in QSOs \cite{DESI2024.II.KP3}. While we do not focus this analysis on the DR1 data, we do discuss some of the Y1 pipeline outputs and validation that the pipeline is working as expected. The relative incompleteness of the DR1 sample makes the \AMTL pipeline a vital piece in the analysis, especially for the creation of realistic mocks \cite{KP3s8-Zhao,KP3s6-Bianchi} and analyses of small-scale clustering.


\section{Mocks}
\label{sec:mocks}

In order to validate the \AMTL method, we use a set of 25 mock realizations processed in a similar manner to the data, reproducing the SV3 processing. They are built from the dedicated  \ \textsc{AbacusSummit}\footnote{https://abacussummit.readthedocs.io/en/latest/} N-body simulations \cite{abacus1,abacus2} and include different ensembles for LRG, ELG and QSO. These are then combined into a single file that mimics the structure of the observation data and that are suitable to be processed through fiber assignment and LSS pipelines.

Mocks are made from Abacus realizations with fiducial cosmology (Planck 2015 $\Lambda$CDM cosmology \cite{planck2015}). For each tracer type (ELG, LRG and QSO) we build cubic mocks of 2 \gpch size each, at different primary redshifts following a HOD modelling \cite{abacusHODELG, abacushod2} tuned to the clustering amplitude of SV3 data. In SV3, we chose the following Abacus snapshots: LRGs at z=0.8, ELGs at z=1.1 and QSOs at z=1.4. 

Once the cubic box mocks are made, we transform them to CutSky mocks, translating from Cartesian coordinates to sky coordinates and including the effect of redshift-space distortions, resulting in galaxy catalogs already in redshift space. These catalogs are resampled to match the same n(z) distribution and footprint as SV3 data and prepared such that we add the columns needed to go through fiber assignment. In addition, we also make random catalogs following the same footprint and redshift distribution as the mock catalogs.

These mock catalogs are then processed through the \AMTL pipeline (describe in \S\ref{sec:altmtl}) and LSS pipeline (described in \S4 in \cite{edr}), creating clustering catalogs analogous to those of the data. Not all the observational effects are modelled, and therefore, we adapt the \AMTL and LSS pipelines to accommodate these differences. 

The only significant difference in the \AMTL pipeline is that we do not pull any information from the redshift-fitting process \cite{SpecPipe} and the effect of redshift failures is ignored through the entire \AMTL loop (fixing \texttt{ZWARN} = 0 in all steps). In summary, we assume that all targets assigned have a good redshift. In future releases we expect to model this effect more robustly, although we expect negligible effect on the clustering signal, since redshift failures will depend on the individual physical properties of the targets and is assumed to be independent of the underlying clustering. 

Finally, the output products of the mock \AMTL method are run through the LSS pipeline to create clustering ready catalogs with the same completeness properties as the data. Together with the mock data, we process the associated randoms. Here we have modified the pipeline to ignore any effect coming from low data quality (assuming a 100\% success rate) and ignoring any angular mask applied to the data. These implied subtle changes in the pipeline at the full catalog stage (see \S4.2.1 in \cite{edr}).

\section{Alternate Merged Target Ledgers}
\label{sec:altmtl}

\subsection{Pairwise Inverse Probability Weighting}

Before describing the Alternate MTLs, we will first describe how they are used in the context of the Pairwise Inverse Probability (PIP) weighting of BP17. 

PIP weighting, as the name suggests, involves weighting pairs of galaxies based on the inverse of the probability that each individual pair would be observed. This probability can theoretically be calculated analytically, but the large number of galaxies observed in DESI ($\mathcal{O}(10^7)$) prohibits this from being practical. Instead of this, we can run Monte Carlo realizations of DESI, varying only those selection properties which are truly random.


When deciding how many realizations to generate for the calculation of PIP weights, we consider that in order for the PIP weights to formally unbiased, there must be no zero probability pairs. While we cannot guarantee zero probability pairs in a finite number of realizations, we show in Table~\ref{tab:ZPOTable} that in SV3 there no zero probability QSOs and BGS\_BRIGHT targets with $\mathcal{O}(10^{-3})\%$ zero probability LRGs and ELGs while in Y1, the fraction of zero probability objects is less than 1\%. 

To mitigate the effect of zero probability pairs, we also employ the angular upweighting of PB17. The combination of PIP weighting with angular upweighting was shown in Bianchi et al 2018 \cite{Bianchi18} as well as in subsequent analyses of data and simulations \cite{Mohammad18, Smith19,Mohammad20} to provide practically good results in the presence of zero probability pairs despite a lack of perfect theoretical rigor.

\subsection{The Merged Target Ledgers}
DESI stores potential targets in a set of files referred to as the merged target lists or ledgers (MTLs; see \S6 of
\cite{SurveyOps}). These MTLs contain information generated and updated by the DESI targeting software \cite{DESITargetSelection} and used by the DESI assignment code \citep{fba}, to place fibers on targets. This information includes a unique \texttt{ TARGETID} field to identify objects, sky coordinates expressed as Right Ascension (\texttt{ RA}) and Declination (\texttt{ DEC}), a \texttt{ PRIORITY} (determined by a target's  class), a randomly determined \texttt{ SUBPRIORITY} (in the range 0 to 1), and bitmasks indicating the different target classes and programs to which an object belongs. A summary of all the parameters in the MTLs is given in table \ref{tab:MTLParams} and a more detailed description is given in \S6 of \cite{SurveyOps}.
\begin{table}
\label{tab:MTLParams}
\caption{List of parameters in a merged target ledger (see also, e.g., \url{https://desidatamodel.readthedocs.io/en/stable/DESI_SURVEYOPS/mtl/sv3/dark/sv3mtl-dark-hp-HPX.html})}
\begin{tabular}{|p{3.75cm}|p{12.5cm}|}
\hline
Parameter Name  &  Parameter Description \\
\hline
\texttt{ TARGETID} & Unique identifier for each DESI target (see \cite{DESITargetSelection}). \\
\texttt{ RA} & Right Ascension in decimal degrees. \\ 
\texttt{ DEC} & Declination in decimal degrees. \\  
\texttt{ REF\_EPOCH} & Reference epoch for stellar targets used to adjust coordinates for proper motion. \\
\texttt{ PARALLAX} & Component of stellar coordinate change due to parallax. \\ 
\texttt{ PMRA/PMDEC} & Component of stellar coordinate change due to true (proper) motion of stars. \\
\texttt{ SV3\_DESI\_TARGET /SV3\_BGS\_TARGET} /\texttt{ SV3\_MWS\_TARGET /SV3\_SCND\_TARGET} & Bitmasks described completely in \texttt{ desitarget.targetmask} (see \cite{DESITargetSelection}) which record which target class(es) a target falls into. \\
\texttt{ PRIORITY} & The priority which fiberassign uses to determine which targets to assign fibers. This is determined by target class and observation status. \\
\texttt{ PRIORITY\_INIT} & The priority for a target that has yet to be observed.  \\
\texttt{ SUBPRIORITY} & A small additional priority which fiberassign uses to reproducibly resolve collisions. This is a random number in the range 0 to 1. \\
\texttt{ OBSCONDITIONS} & Bitmask which represents the observing conditions (bright, dark or backup) in which an object can be targeted (see \cite{SurveyOps}). \\
\texttt{ NUMOBS\_INIT} & The number of observations required for a target that has yet to be observed. \\
\texttt{ NUMOBS\_MORE} & The number of successful observations until the target is considered "done". \\
\texttt{ NUMOBS} & The current number of observations of the target. \\
\texttt{ Z} & The redshift of the target. \\
\texttt{ ZWARN} & Bitmask which records \texttt{ Redrock} warning flags for the redshift determination. \\
\texttt{ ZTILEID} & The ID(s) of the tiles on which a target obtained a good observation. \\ 
\texttt{ TARGET\_STATE} & A human-readable representation of the current state of the target factoring in its various target classes and current status. \\ 
\texttt{ TIMESTAMP} & The timestamp at which the MTL entry was made. \\
\texttt{ VERSION} & The version of the \texttt{ desitarget} code which was used to make the MTL update. \\
\hline
\end{tabular}
\end{table}



When an observation is performed, each target for which good data was obtained has an entry added to the end of ledger that records the results of the observation. A highly simplified example of the initial ledger is presented in table \ref{tab:initSimpleMTL} and the result of updating the states in the simplified initial ledger is presented in table \ref{tab:updatedSimpleMTL}. As shown in tables \ref{tab:initSimpleMTL} and \ref{tab:updatedSimpleMTL}, the updates made to a target after a successful observation include an increment in \texttt{ NUMOBS}, a drop in \texttt{ PRIORITY}, an associated change in \texttt{ TARGET\_STATE}, and updates to the \texttt{ Z} and \texttt{ ZWARN} fields based on the output from the DESI spectroscopic pipeline \cite{SpecPipe} and redshift-fitting software.

\begin{table}
\centering
\caption{Simplified initial merged target ledger}
\setlength\tabcolsep{2pt}
\begin{tabular}{cccccc}
\label{tab:initSimpleMTL}
TARGETID  &  PRIORITY & NUMOBS & Z & ZWARN & TARGET\_STATE\\
\hline
1 & 103200 & 0 & -1.0 & -1 & LRG$\|$UNOBS \\
2 & 103100 & 0 & -1.0 & -1 & ELG$\_$HIP$\|$UNOBS \\
3 & 103400 & 0 & -1.0 & -1 & QSO$\|$UNOBS

\end{tabular}
\end{table}

\begin{table}
\centering
\caption{Simplified updated merged target ledger}
\setlength\tabcolsep{2pt}
\begin{tabular}{cccccc}
\label{tab:updatedSimpleMTL}
TARGETID  &  PRIORITY & NUMOBS & Z & ZWARN & TARGET\_STATE\\
\hline
1 & 103200 & 0 & -1.0 & -1 & LRG$\|$UNOBS \\
2 & 103100 & 0 & -1.0 & -1 & ELG$\_$HIP$\|$UNOBS \\
3 & 103400 & 0 & -1.0 & -1 & QSO$\|$UNOBS \\
1 & 2 & 1 & 0.684 & 0 & LRG$\|$DONE \\
2 & 103100 & 1 & 1.45 & 4 & ELG$\_$HIP$\|$MORE\_ZWARN \\
3 & 103350 & 1 & 1.9 & 0 & QSO$\|$MORE\_MIDZQSO
\end{tabular}
\end{table}

The updating of MTLs in-between repeat observations of the same patches of sky combined with the resulting drop in priority for a successfully observed targets highlights the utility of the \AMTL method. Crucially, a single change in which a target was observed on the first pass through a region can cause cascading effects due to fiber placement on subsequent tiles. 

\subsection{Generating the Alternate Merged Target Ledgers}
\label{sec:constructAMTL}
Having described the regular MTLs, we will now describe how we generate the alternate MTLs.

The primary difference between the regular and alternate MTLs is the randomization of the subpriority field. The other major difference is, for both SV3 and main survey, a reproduction of the BGS promotion of 20\% of BGS\_FAINT targets to the higher priority class of BGS\_BRIGHT, and for the main survey only, a reproduction of the ELG promotion of 10\% of all ELG targets to the higher priority ELG\_HIP class.

To begin the alternate MTL generation process, the survey MTLs are loaded and then truncated so that only the initial entries are maintained. For the initial entries, the SUBPRIORITY field is replaced with new random numbers from the basic numpy uniform random number generator. The random number generation is seeded based on the healpixel number of the initial MTL, the realization number, and a user provided random seed. The SV3 MTLs were randomized with a version of the code which did not add in the healpixel number to the random seed. This randomization is still valid, but we include this information, as well as a tagged version of the code as it was for the SV3 MTLs\footnote{https://github.com/desihub/LSS/tree/v1.0.0-EDA} to allow for reproduction. The promotion of the BGS targets is a multiple step process where first, the initially promoted 20\% are demoted back to the standard BGS\_FAINT priority and then a desitarget convenience function (\texttt{random\_fraction\_of\_trues}) is used to select 20\% of the total BGS\_FAINT targets to promote to BGS\_FAINT\_HIP. The promotion of ELG targets is similar with two additional considerations. First, targets are being drawn from two starting classes: ELG\_LOP and ELG\_VLO and must be returned to their resepective classes. Second, not all demoted objects will see a drop in priority since it is not forbidden to have an object be targeted as both an ELG and an LRG (same priority as ELG\_HIP) or as both an ELG and QSO (higher priority than ELG\_HIP).

\begin{figure*}
    \label{fig:ZPOTrendSV3}
    \centering
    \includegraphics[scale=0.4]{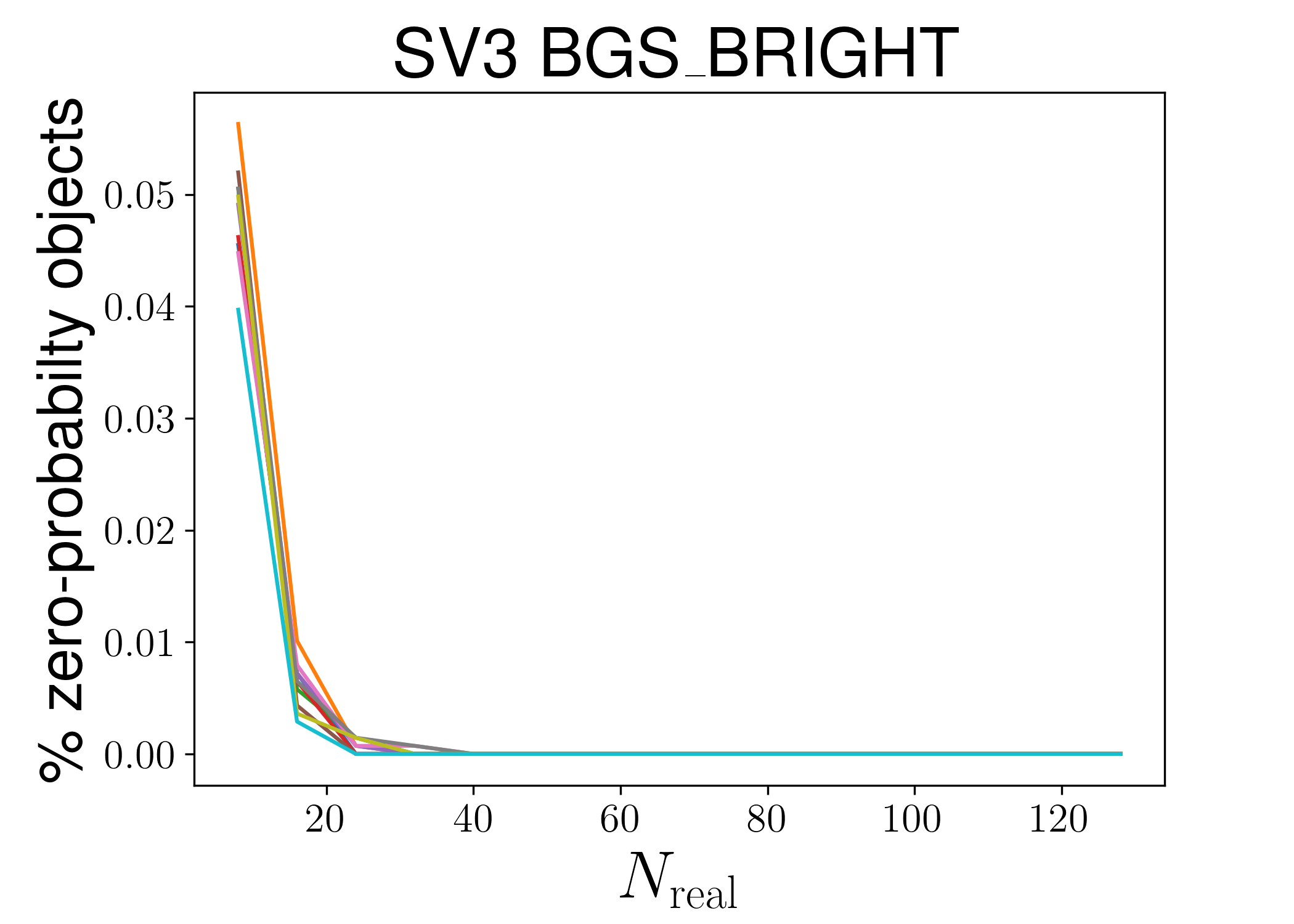}
    \includegraphics[scale=0.4]{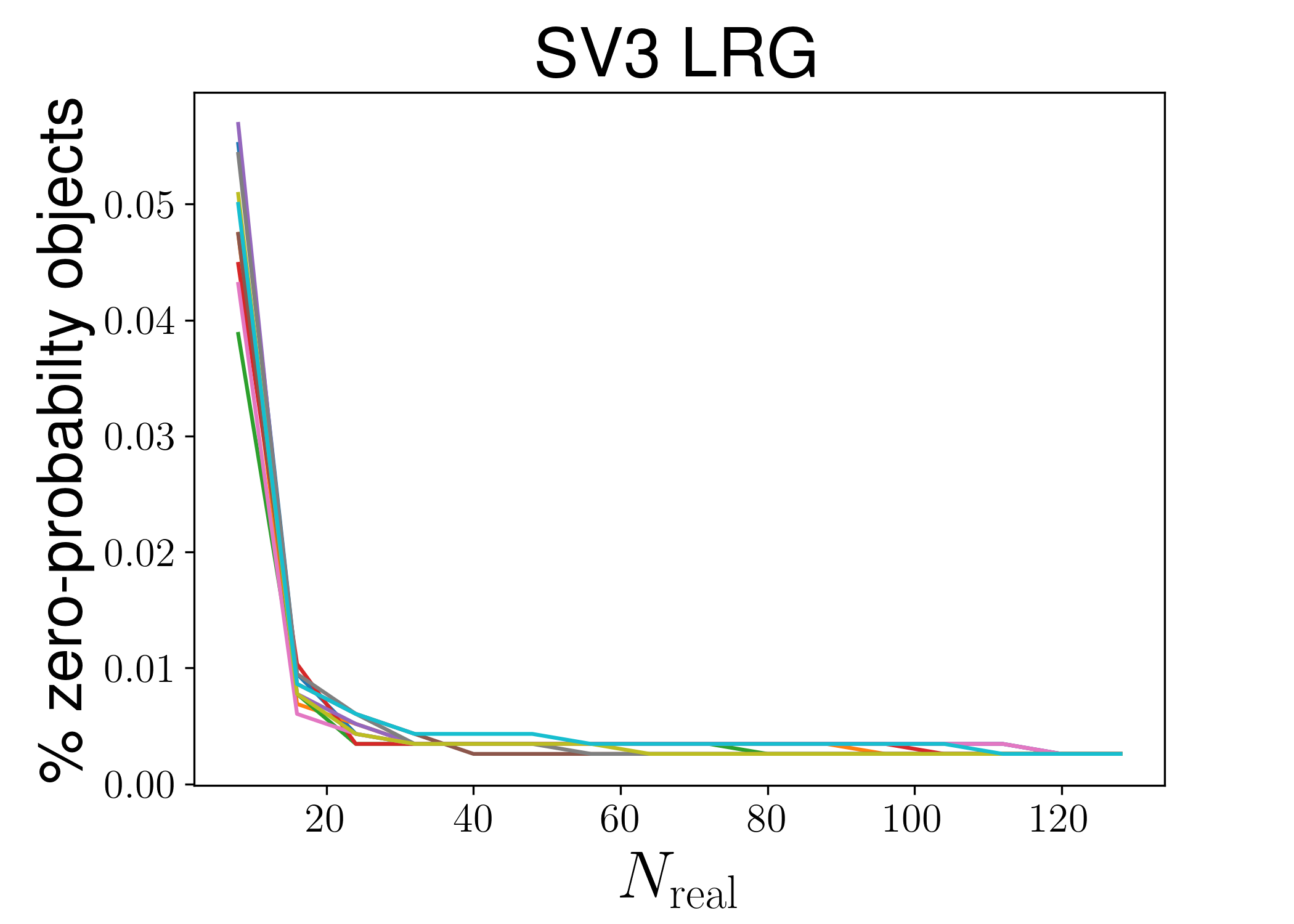}
    \includegraphics[scale=0.4]{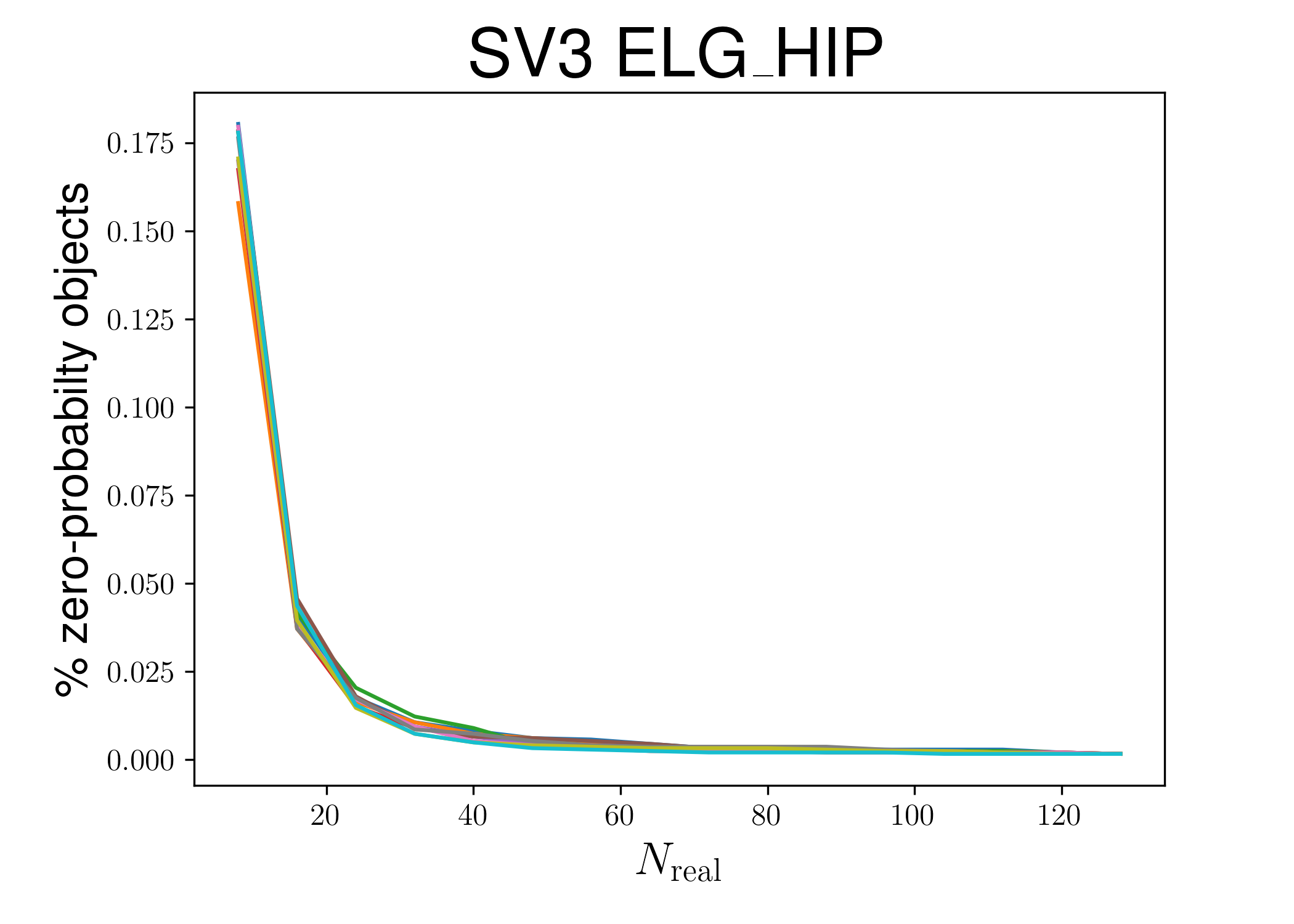}
    \includegraphics[scale=0.4]{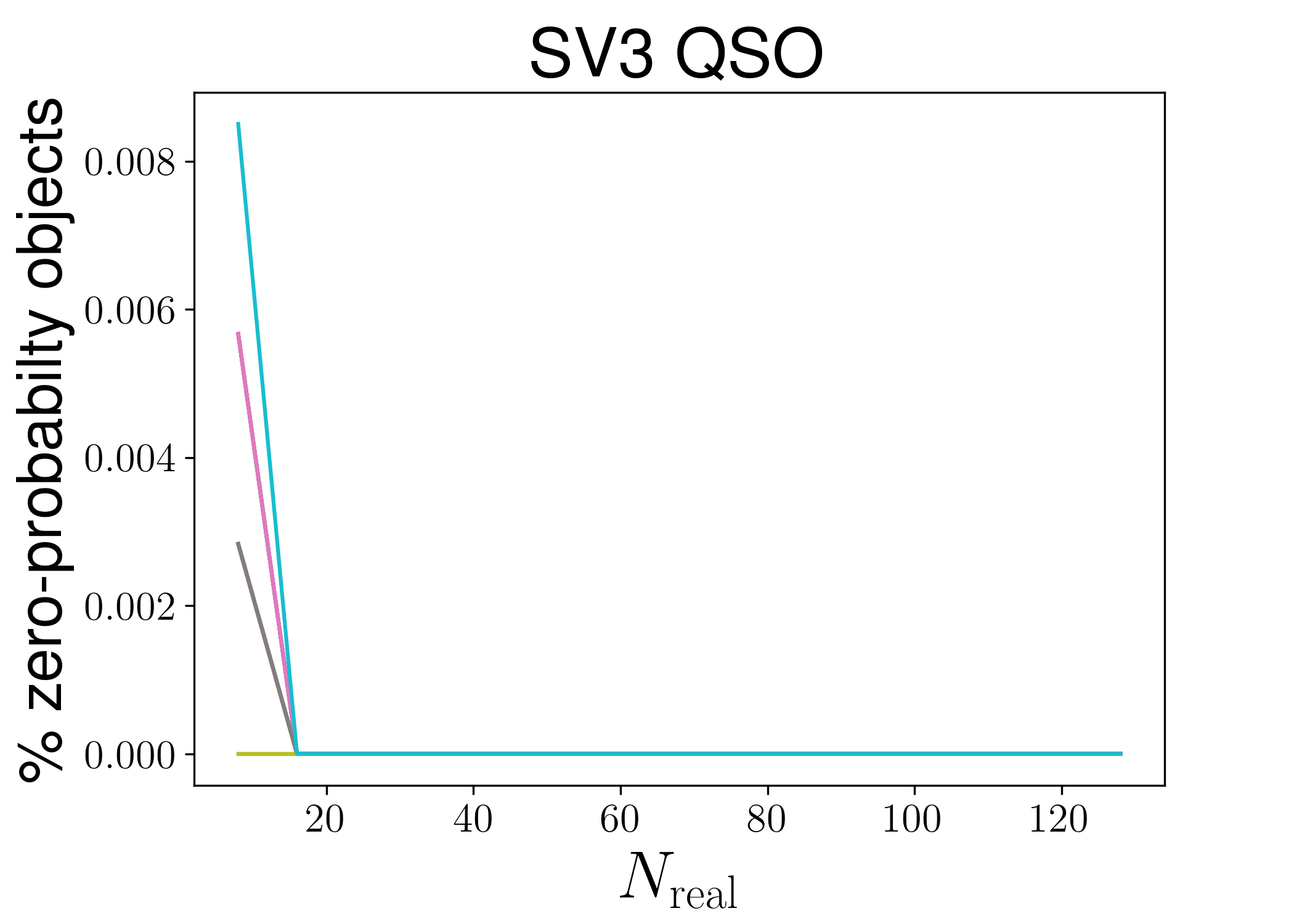}
    \caption{The percentage of objects not observed in any alternate realization of SV3 as the number of realizations increases from 8 to 128. Each line within a plot represents a random reordering of the 128 realizations to confirm that there is no dependence of the convergence to zero on the (random) order of the realizations. }
\end{figure*}

\begin{figure*}
    \label{fig:ZPOTrendY1}
    \centering
    \includegraphics[scale=0.4]{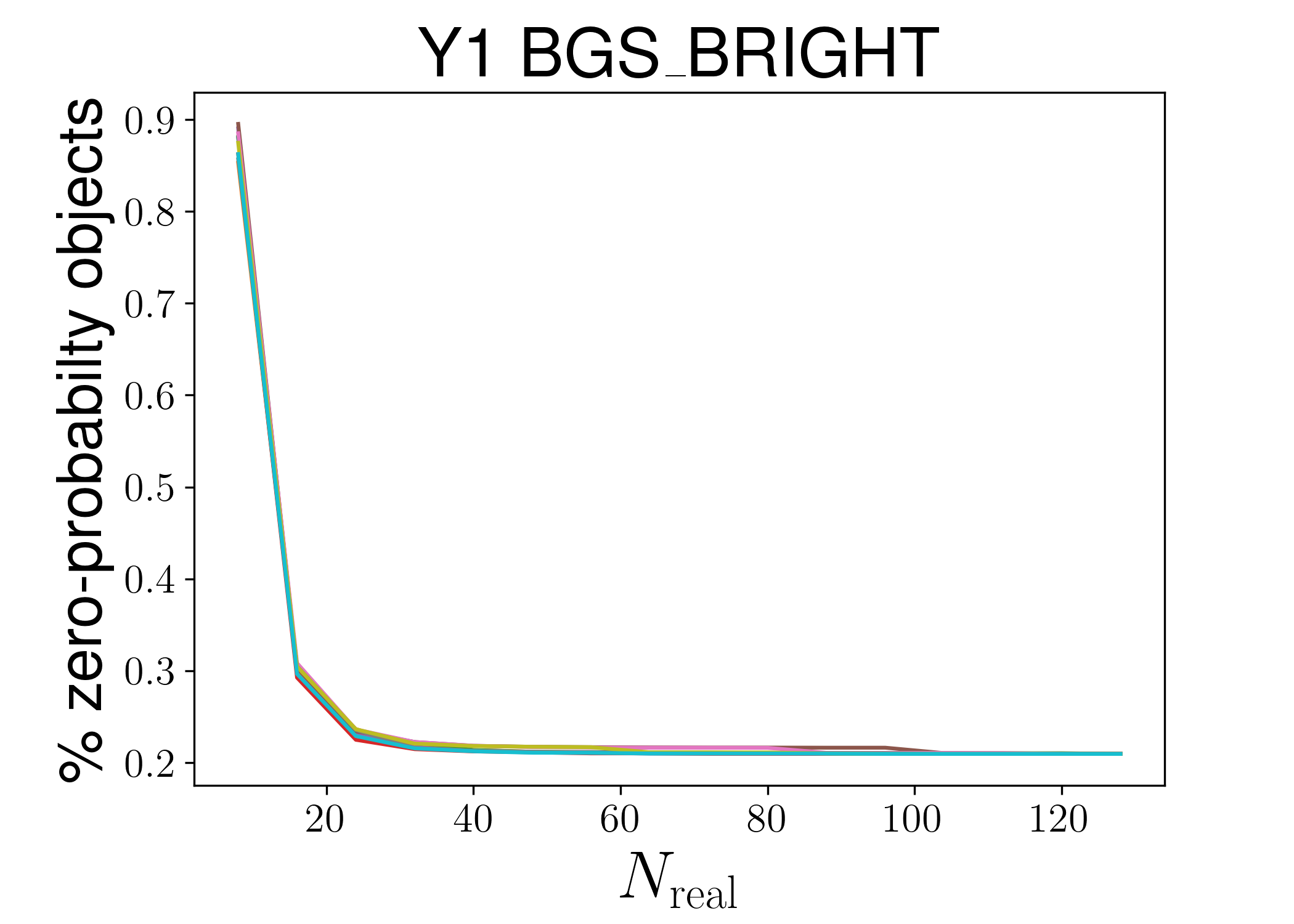}
    \includegraphics[scale=0.4]{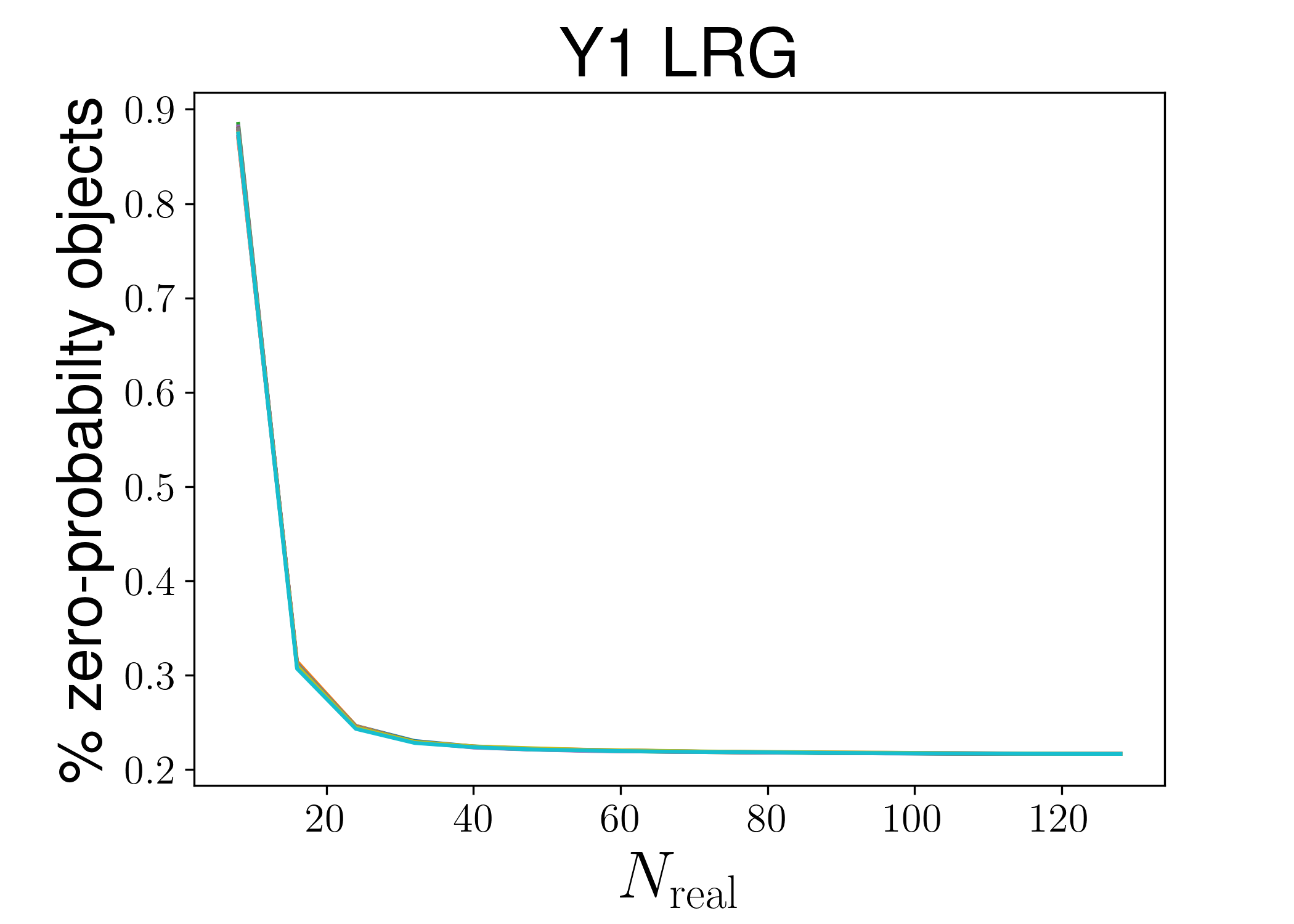}
    \includegraphics[scale=0.4]{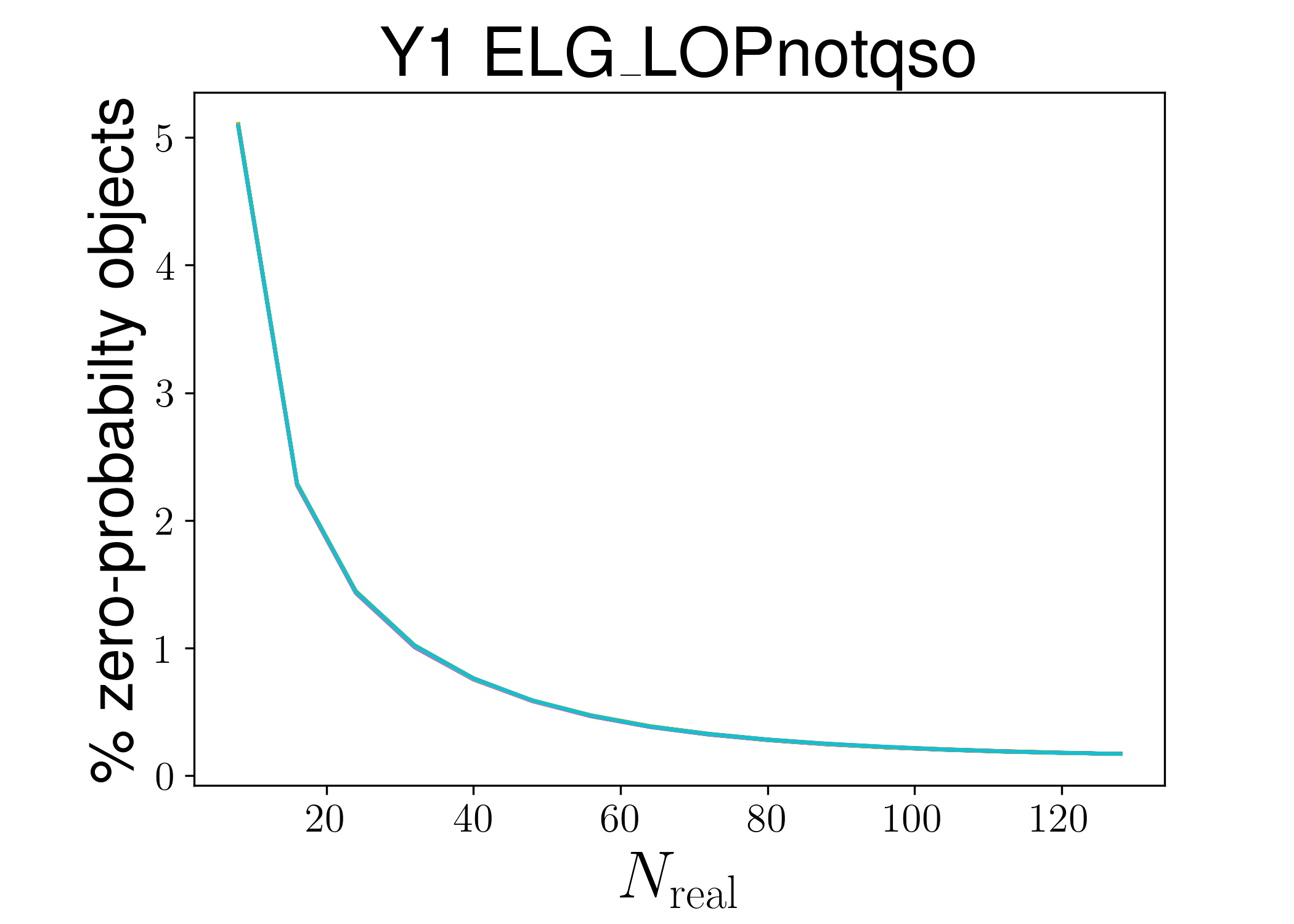}
    \includegraphics[scale=0.4]{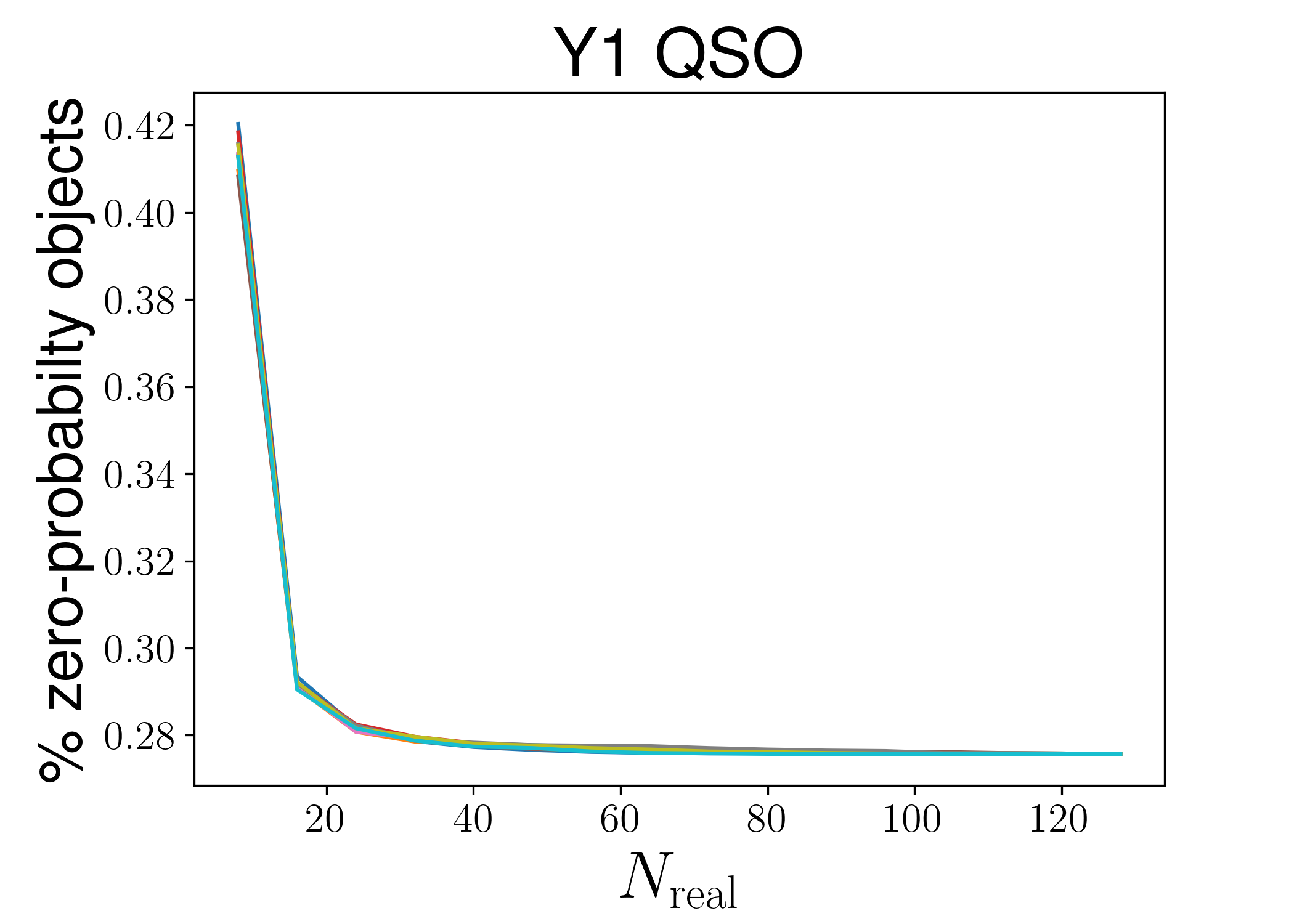}
    \caption{The same as Figure \ref{fig:ZPOTrendSV3} but for Y1 targets. Note that the y-axis floor here is not zero. }
\end{figure*}

\begin{table}
\centering
\caption{Percentage of zero probability objects over 128 realizations of SV3 and Y1}
\setlength\tabcolsep{2pt}
\begin{tabular}{ccc}
\label{tab:ZPOTable}
Tracer type  & \% of Zero Probability & \% of Zero Probability \\
 & Objects in SV3 & Objects in Y1 \\
\hline
BGS\_BRIGHT & 0 & 0.21 \\
LRG & 3E-3 & 0.22 \\
ELG & 3E-3 & 0.17 \\
QSO & 0 & 0.28 

\end{tabular}
\end{table}
This procedure is done in parallel for up to 128 alternate MTLs per node on the Perlmutter supercomputer at NERSC.\footnote{www.nersc.gov/perlmutter} For the SV3 alternate MTLs, we have chosen to only have the 128 realizations of alternate MTLs for the data and 256 for the mocks. This choice was made from a consideration of the number of zero probability objects that result from a given number of realizations as shown in figure \ref{fig:ZPOTrendSV3} for SV3 and \ref{fig:ZPOTrendY1} for Y1. Each panel of both figures shows the number of zero probability objects from each target type (from left to right: BGS\_BRIGHT, LRG, ELGs with QSO targets removed (ELG\_notqso), and QSO) as the number of realizations increases from 8 to 128. The multiple lines in each figure are random reorderings of the 128 realizations to confirm that there is no dependence of the convergence on the (random) ordering of realizations. The exact percentages of zero probability objects for SV3 data and Y1 data for the chosen number of 128 realizations is given in Table~\ref{tab:ZPOTable}. A zero probability object is a target  that is never observed over a given number of realizations. BP17 states that their method of weighting is not completely valid where there are zero probability pairs. While we have essentially eliminated zero probability objects in SV3, the trend in Figure~\ref{fig:ZPOTrendY1} shows that it would be prohibitive to run enough realizations to eliminate all zero probability objects in Y1. However, we will show that the perecentages of zero probability objects shown in table \ref{tab:ZPOTable} from 128 realizations will sufficiently mitigate the problem. 

\subsection{The Observation Loop}
\label{sec:ObsLoop}
The core of the utility of the \AMTL method as compared to other implementations of PIP weighting is the ability to update the alternate MTLs using real observation information. In order to do this, we create an observation loop that follows the same ordering as the survey data processing and observation. 

Following the creation of the initial randomized ledgers, the SV3 \AMTL loop would use the \texttt{ desitarget.mtl} function \texttt{ tiles\_to\_be\_processed} to check the list of observed SV3 tiles from the data against the list of already processed tiles for each realization. Then the set of tiles corresponding to the earliest remaining observation date were selected. All tiles which were observed on that date were put through an observation loop designed to replicate the SV3 observing process. 

The observation loop begins by running fiberassign \citep{fba} using the alternate ledgers as the primary target ledgers. Other than the primary target ledgers, this procedure uses the same configuration as the corresponding fiber assignment run in the real survey including the list of sky fibers, calibration standards, secondary targets, and the hardware configuration at the exact time at which fiberassign was run for the main survey. This produces a set of files which map the alternate targets to DESI fibers for the alternate observations. 

In order to accurately track the completeness effects due to improperly positioned fibers, fiber positioning system failures, and other focal plane-level effects, we map the observation result from each fiber in the original survey to the same fiber in the alternate realization. For example, if fiber 1 were pointed at target 1 in tile 1 of the real survey and fiber 1 were pointed at target 2 in tile 1 of an alternate realization, target 2 would would receive the observation result of target 1 in that alternate realization. Inclusion of the observational results from the survey means any real-time positioning failures are automatically incorporated. The survey pipeline marks observations as bad if they are come from fibers with flags indicating bad hardware status. 

This can, and does, lead to the assignment of observations of one target class to observations of another target class. However, since the primary consideration is whether or not a target has been placed on a fiber which received a good observation rather than whether it received a successful redshift, this approach is still mostly valid. For the case of QSOs where the value of the redshift determines the number of future observations this is not exactly correct. For this purpose we are developing a method to assign results from a "twin fiber" which is nearby the true fiber and matches in as many properties as possible, primarily the target class. For the full five year sample, we also plan to substitute QSO redshifts from original observations with the final survey redshift results. This will be possible since QSOs are the highest priority target class and will be nearly fully complete by the end of the survey. In this analysis, we will note that the fraction of observations whose primary target information changes is 6.7 \% (13.0\%) of all SV3 (Y1) dark time observations and 13.0 \% (31.8\%) of SV3 (Y1) dark time observations which result in a change in target. 

The main survey data added an additional complication to the observation loop, the addition of reprocessed tiles. A reprocessing occurs when the results of the daily pipeline are modified due to (e.g.) a run of the offline pipeline discovering that a "good`` redshift was not actually a good fit. Since reprocessings can change the state of already observed targets and they occur independently of the observations themselves, reprocessings necessitated a separation of the alternate fiber assignment step and the alternate MTL update step where we check for reprocessings of tiles. This also resulted in the storage of the maps between the real and alternate fibers from the regular observation loops so that the reprocessed redshift catalogs/ledgers can be modified in the same way as the original ones.

\subsection{Outputs of the pipeline}

The \AMTL pipeline produces two primary outputs. First is the collection of N (for SV3 and Y1, N = 128 ) realizations of MTLs which includes information on every (alternate) observation of every target. The columns for these MTLs are the same as in the real survey MTLs, which you can find in table \ref{tab:MTLParams}. 

The second output is a bitweight file with three columns, it can be attached to the LSS catalogs (and is indeed already available for the EDR catalogs.\footnote{https://data.desi.lbl.gov/doc/} \cite{edr})
First is the unique TARGETID which is necessary to match the information to other catalogs. Second is two columns with information on how frequently each target was observed over the 128 realizations. One of those two columns is the probability that the individual target would be observed over those 128 realizations, PROB\_OBS. PROB\_OBS is calculated as the number of realizations in which a target was observed divided by the total number of realizations.  The other remaining column is the bitweights, a shorthand encoding of whether each target was observed in each realization as first outlined in BP17. Bitweights are stored as an ordered list of 64 bit integers where each bit is set to 1 if a target was observed (NUMOBS $\geq$ 1) in a given realization and set to 0 if it was not observed in that realization.\footnote{In order to manipulate these bitweights, we provide the functions within desihub's \texttt{ LSS.bitweights} module}

\subsection{Validation of the pipeline}

Validation of the pipeline has been a multiple step process. Prior to running the full set of 128 realizations, we have run a single realization with the part of the pipeline which randomizes the subpriorities turned off and confirmed that this version of the pipeline is able to reproduce both SV3 and Y1 exactly. We have also validated that the pipeline produces a statistically equivalent set of promotions to ELG\_HIP and BGS\_BRIGHT out of the initial randomized alternate ledgers.

In \S \ref{sec:results} we show clustering measurments validating the PIP weights generated from the pipeline on the mocks described in \S \ref{sec:mocks}.

\section{Clustering Measurements}
\label{sec:TPCFCode}

We will measure clustering based on pair-counts as a function of the redshift-space transverse, $s_{\perp}$, and radial, $s_{||}$ separation, measured in co-moving $h^{-1}$Mpc using the fiducial cosmology. The 2-point correlation function probes the excess probability to find in a galaxy sample two galaxies at a given separation. We use the so-called Landy-Szalay estimator~\cite{LandySzalay}
\begin{equation}
\widehat{\xi}(s_{\perp}, s_{||}) = \frac{DD(s_{\perp}, s_{||}) - DR(s_{\perp}, s_{||}) - RD(s_{\perp}, s_{||}) + RR(s_{\perp}, s_{||})}{RR(s_{\perp}, s_{||})}
\label{eq:ls_estimator}
\end{equation}

where $XY(s_{\perp}, s_{||})$ correspond to the weighted number of pairs of objects $X$ and $Y$ in a $s_{\perp}, s_{||}$ bin divided by the total weighted number of pairs. Specifically, $DD(s_{\perp}, s_{||})$ is the weighted number of data (galaxy and quasar) pairs and $RR(s_{\perp}, s_{||})$ is the weighted number of pairs of objects in the random catalog which samples the selection function. For the autocorrelation (single-tracer) measurements performed in our analysis, by symmetry $DR(s_{\perp}, s_{||}) = RD(s_{\perp}, s_{||})$. For standard estimates pair weights are the product of the total individual weights of the two objects in the pair $w_{1}$ and $w_{2}$, themselves obtained as the product of systematic correction weights and FKP weights.  We use 48 logarithmically spaced $s_{\perp}$ bins between 0.01 and 100 $\mathrm{Mpc}/h$ and 40 linearly spaced $s_{||}$-bins between 0 and $40\; \mathrm{Mpc}/h$. 

To optimize computing time we use the technique of Keih{\"a}nen et al 2019~\cite{SplitRandom}, which consists in summing $DR$, $RD$ and $RR$ pair counts obtained using several random catalogs of a size approximately matching that of the data catalog. In practice, we use the maximum number available for the SV3 LSS catalogs, which is 18 and corresponds to 28, 67, 220, and 54$\times$ the data number for ELGs, LRGs, QSOs, and BGS\_BRIGHT respectively.

In this work, we will show the results of the projected correlation function $w_p(r_p)$, with $r_p \equiv s_{\perp}$ and
\begin{equation}
    w_p(r_p) = 2\int_0^{40} ds_{||}\xi(r_p, s_{||})
\end{equation}

the projected correlation function $w_p$ integrates the full two point correlation function $\xi$ over a range in the line-of-sight distance spanning 40 \mpch. This distance is chosen to sufficiently remove all rsd effects while not introducing significant noise.

Correlation function measurements are performed with \textsc{pycorr}\footnote{\url{https://github.com/cosmodesi/pycorr}} which wraps a version of the \textsc{Corrfunc} package~\cite{Corrfunc1, CorrFunc2} modified to also run on GPU and support alternative lines-of-sight (first-point, end-point) and weight definitions (PIP and angular upweights) and the $\theta$-cut, along with jackknife error estimation.

\subsection{Weights}
\label{sec:ClusteringWeights}

The set of weights applied in the DESI SV3 TPCFs include the "default`` (IIP) weights, FKP weights, angular upweights, and bitwise (PIP) weights. The PIP and IIP weights are never applied simultaneously since they are derived from the same set of {\AMTLnospace}s and correct for the same incompleteness. 

The ''default`` IIP weights are calculated from the bitweights output from the \AMTL pipeline with the 'efficient estimator` of Bianchi and Verde 2020 \cite{BianchiVerde20}. 

\begin{equation}
w_{i,\rm IIP} = \frac{N_{\rm real} + 1}{ ({\rm popcnt} \: w^{(b)}_i) + 1}
\end{equation}

FKP weights were derived in Feldman, Kaiser, and Peacock 1994\cite{FKPWeights} as a variance-minimizing weight for a given scale of a power spectrum. These FKP weights are given by

\begin{equation}
    w_{FKP}(z) = \frac{1}{1 + n(z)P_{0}}
\end{equation}

where $n(z)$ is the number density of galaxies at redshift $z$, and $P_{0}$ is the clustering power at the scale in the power spectrum at which we desire to minimize the variance. In DESI SV3, we chose the following $P_{0}$ 
 values for each tracer: 

LRG: 10000
ELG: 4000
QSO: 6000
BGS: 7000

Angular upweighting, as first outlined in PB17 \cite{PercivalAndBianchi} and applied in Mohammad et al 2018 \cite{Mohammad18} and Mohammad et al 2020\cite{Mohammad20}, is derived from the angular clustering of the parent catalog of all potential targets of a tracer type and the angular clustering of the targets which have been observed. These pairwise weights are generated separately for the Data-Data pairs and the Data-Random pairs with the bitweights factoring into both sets of counts as shown below.

\begin{equation}
\label{eq:ANGUpDD}
 w^{\rm DD}_{\rm ANG} = \frac{DD^{\rm par}(\theta)}{DD^{\rm fib}_{\rm PIP}(\theta)}
\end{equation}

\begin{equation}
\label{eq:ANGUpDR}
 w^{\rm DR}_{\rm ANG} = \frac{DR^{\rm par}(\theta)}{DR^{\rm fib}_{\rm IIP}(\theta)}
\end{equation}

In equations \ref{eq:ANGUpDD} and \ref{eq:ANGUpDR}, "par`` and "fib`` refer to the parent sample and fiber-assigned samples respectively. The fiber-assigned Data-Data pairs are weighted by the inverse of the probability that the pair would be observed (PIP). Similarly, the fiber-assigned Data-Random pairs are weighted by the inverse of the probability that the data galaxy would be observed (IIP).

Finally, the bitwise PIP weights are calculated using the efficient estimator from Bianchi and Verde 2020 \cite{BianchiVerde20}:

\begin{equation}
    w_{ij} = \frac{N_{\rm real} + 1}{ {\rm popcnt}\: {w^{(b)}_i {\rm and} \: w^{(b)}_j} + 1 }
\end{equation}

where $w_{ij}$ is the weighting applied to the pair of galaxies $i$ and $j$, $N_{\rm real}$ is the number of \AMTL realizations, and popcnt $w^{(b)}_{i} and w^{(b)}_j$ is the count of the bitwise "and`` operation between the bitweights of galaxies i and j. In the case of PIP weighting, the count of DD pairs is also normalized as in Equation 22 of Bianchi and Verde 2020 \cite{BianchiVerde20}. 

These weights are provided to pycorr by multiplicatively combining all non-bitwise weights and then adding them to a (list of) bitweight(s). 









\begin{figure*}
    \centering 
    \includegraphics[width=\columnwidth]{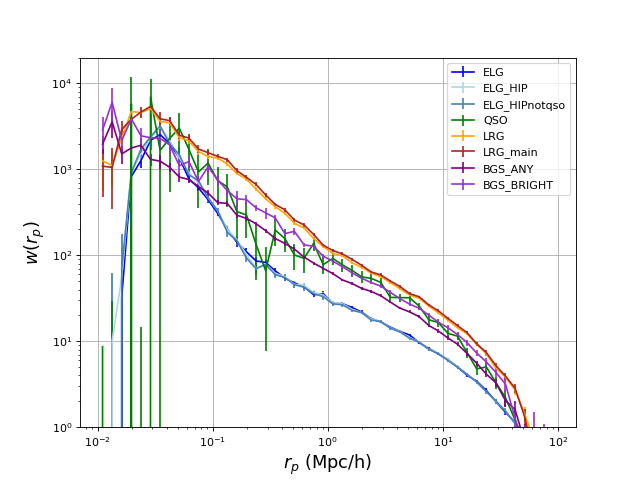}
    \caption{The projected clustering of all tracer types that have a `clustering' catalog produced from DESI SV3 data.}
    \label{fig:wpall}
\end{figure*}

Figure~\ref{fig:wpall}  shows the projected clustering results using the above methods for all tracer types which have a 'clustering` catalog produced from DESI SV3 Data. 

\begin{figure*}
    \centering 
    \includegraphics[width=0.5\columnwidth]{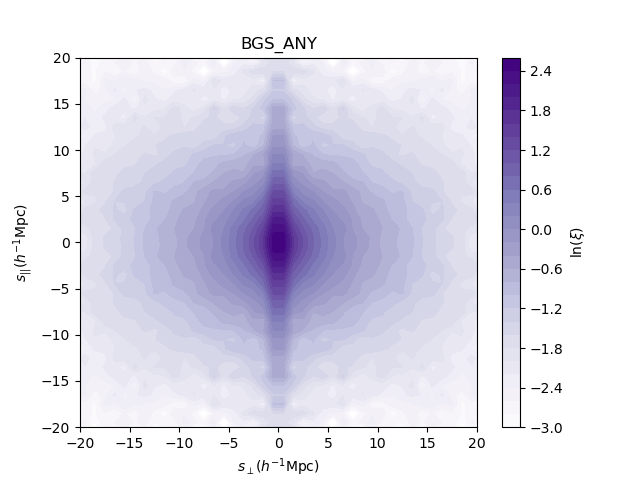}
    \includegraphics[width=0.5\columnwidth]{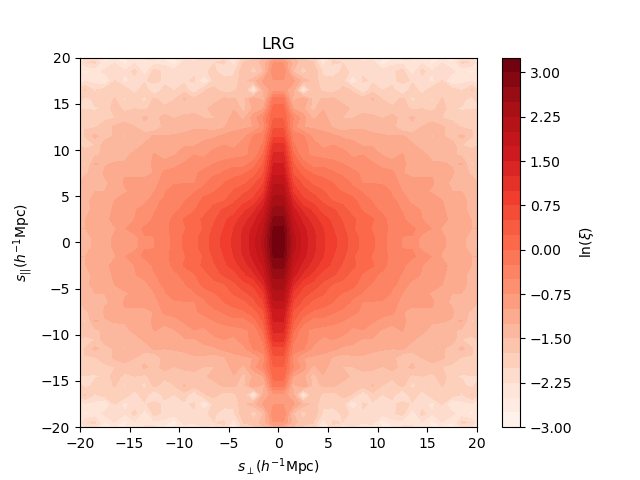}
    \includegraphics[width=0.5\columnwidth]{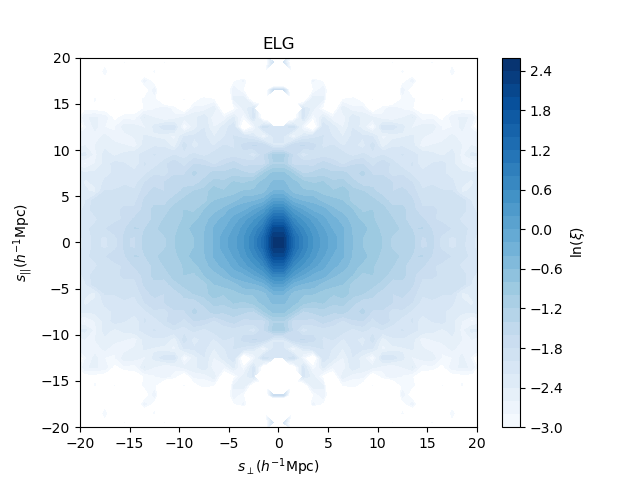}
    \includegraphics[width=0.5\columnwidth]{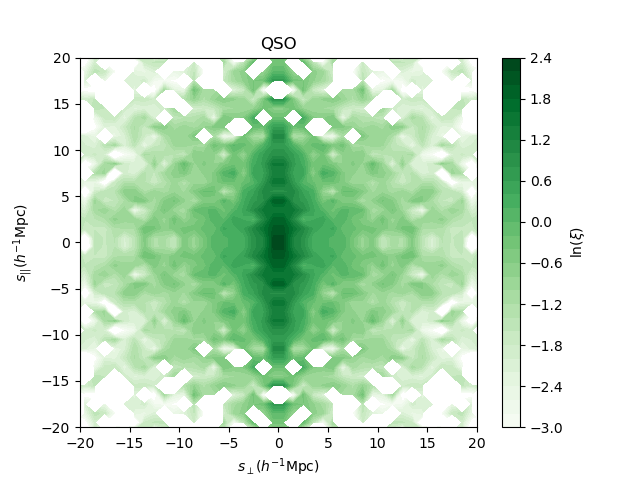}
    \caption{Contour plots showing the clustering amplitude as a function of $s_{\perp}$ and $s_{||}$.}
    \label{fig:xiContour}
\end{figure*}

Figure~\ref{fig:xiContour} shows the clustering amplitude of four tracers (BGS\_ANY, LRG, ELG, and QSO) as a function of their separation in bins of distance parallel and perpendicular to the line of sight ( $s_{||}$ and $s_{\perp}$ respectively).

\section{Results}
\label{sec:results}
In this section we display results from the first two runs of the \AMTL pipeline, one run covering SV3, and the other run covering Y1. Statistics which we will show include a comparison of the number of targets in each class observed in the data and averaged over the \AMTL runs, the estimation of completion from the \AMTL pipeline, the difference in the completeness weights between the \AMTL pipeline and methods which bootstrap completeness estimation from only the data, and finally the effects of weighting using PIP weights from this pipeline and other weighting methods on clustering statistics. 

\subsection{Alternate MTL statistics data}
\subsubsection{Comparison of number of observed targets by class}
Since the \AMTL pipeline functions by assigning real observation results from the DESI survey to alternate targets, we should end up with similar number of total observed targets between the average \AMTL realization and in the data for the major target classes. 

In table \ref{tab:NOBSTable} we show the number of observed targets in the classes BGS\_BRIGHT, LRG, ELG, and QSO in the SV3 Data and the Y1 Data as well as the sum of the PROB\_OBS values for each of those target classes in the 128 realizations of the AMTLs. 

\begin{table}
\centering
\caption{Number of targets of each class observed in data and averaged over 128 \AMTL realizations of SV3 and Y1}
\setlength\tabcolsep{2pt}
\begin{tabular}{ccccccc}
\label{tab:NOBSTable}
Tracer type  & \# Observed & Sum of PROB\_OBS & \%  & \# Observed  & Sum of PROB\_OBS & \%   \\
& in SV3 & in SV3 & difference & in Y1 & in Y1 & difference \\
\hline
BGS\_BRIGHT &   147721 & 148956 & 0.8 & 4234056 & 4274506  & 1.0 \\
LRG & 127020 & 128506 & 1.2 & 2477439 & 2503030 & 1.0 \\
ELG & 354398 & 358432 & 1.1 & 3985829 & 4021208 & 0.9 \\
QSO & 57060 & 57341 & 0.5 & 2036499 & 2051294  & 0.72

\end{tabular}
\end{table}

Table \ref{tab:NOBSTable} reveals a small, but consistent surplus in the number of observations in the AMTLs vs the data. This difference is also present and of similar magnitude in the mocks. Since the mock "observation" sample is produced in the same way as a random \AMTL realization (see table~\ref{tab:NOBSTableMock}), we believe this is reasonable behavior.

\subsubsection{SV3 completeness estimation from \AMTL}
\begin{figure*}
\centering
\label{fig:Rosette16PROBOBS}
\includegraphics[scale=0.28]{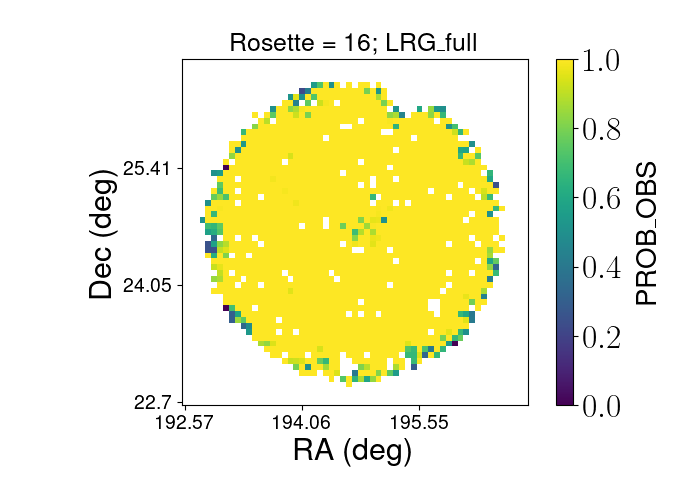}
\includegraphics[scale=0.28]{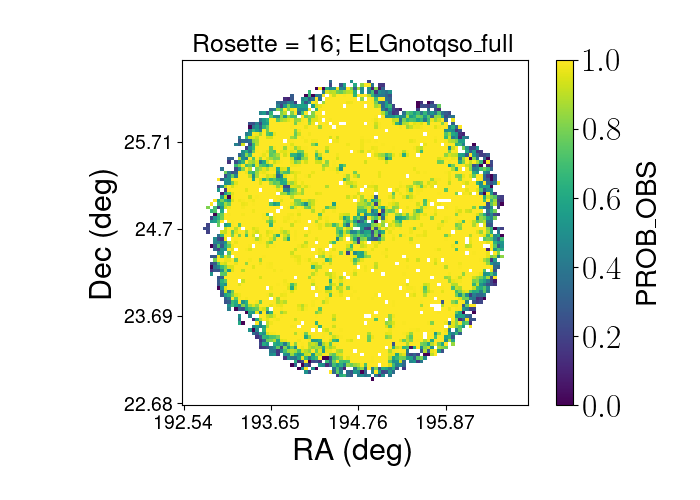}
\includegraphics[scale=0.28]{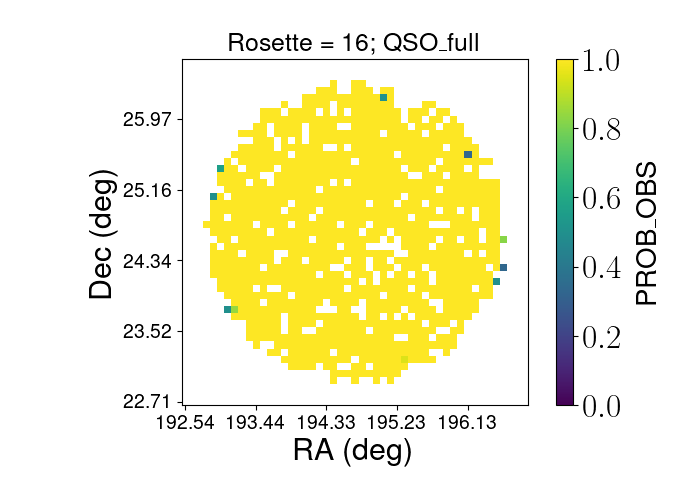}
\caption{2D histogram showing the mean value of PROB\_OBS in bins of RA and Dec within Rosette 16 of SV3 for (from left to right) LRGs, ELGs with QSO targets removed, and QSOs.}
\end{figure*}

Figure~\ref{fig:Rosette16PROBOBS} shows the completeness of one of the SV3 rosettes (rosette 16; rosettes are described in \S \ref{sec:survey} and \cite{SVOverview}) as measured by the average PROB\_OBS for targets of (left to right) LRGs, ELGs with QSO targets removed (ELG\_notqso), and QSOs in RA/Dec bins within the rosette. The binning is varied in width based on the number of targets of each type with ELGs having the most targets and QSOs having the fewest. The tiling pattern of the rosette leads to the high, nearly 100\% completeness in the intermediate parts of the rosette between 0.2 and 1.4 degrees from the center. Regions near the center of the rosette are subject to hardware failures more frequently since they are always covered by the same part of the focal plane and regions towards the exterior of the rosette are only tiled a few times. 

In order to study the potential for the \AMTLnospace-derived bitweights to impact clustering relative to weighting derived from other measures of incompleteness, we will compare the \AMTLnospace-derived observation probabilities with those other measures of incompleteness, focusing our comparisons on the most central and furthest exterior regions where completeness is not 100\%. 

\begin{figure*}
\centering
\label{fig:Rosette16TargetNumDens}

\includegraphics[scale=0.25]{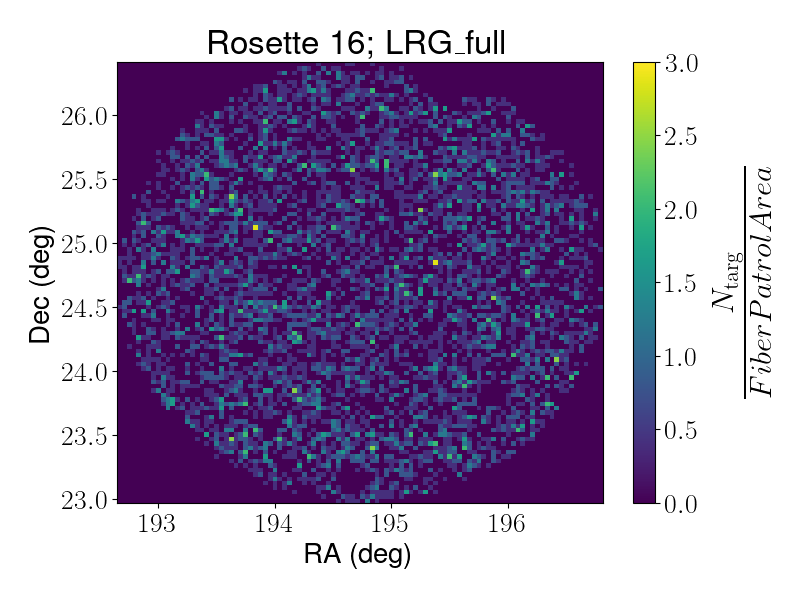}
\includegraphics[scale=0.25]{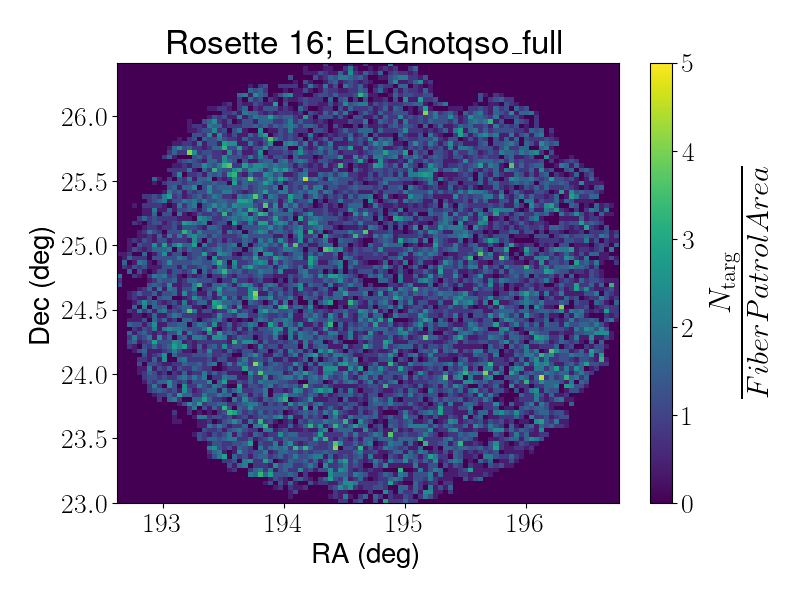}
\includegraphics[scale=0.25]{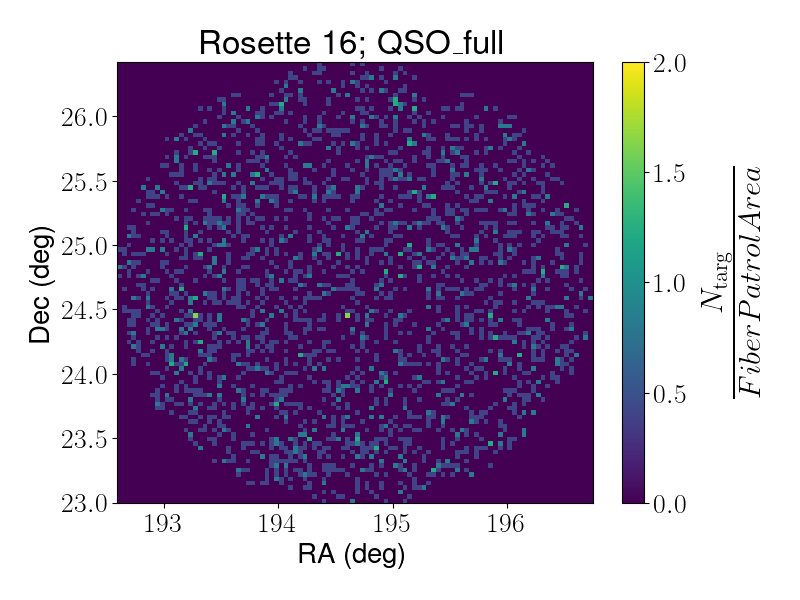}
\caption{2D histogram showing the areal number density (in units of targets per fiber patrol area) of targets in bins of RA and Dec within Rosette 16 of SV3 for (from left to right) LRGs, ELGs with QSO targets removed, and  QSOs.}
\end{figure*}

Figure~\ref{fig:Rosette16TargetNumDens} shows the number density of targets within rosette 16 for each of the same groups of targets in the same order as Figure~\ref{fig:Rosette16PROBOBS}. This is shown in units of number of targets per fiber patrol radius to better express the likelihood that a given bin in RA/Dec will experience fiber collisions. Focusing specifically on the upper left portion of the "ELGnotqso\_full", we can see a correlation between regions of high number density and additional areas of low completeness where we can display the impact of the \AMTL incompleteness weights. 

\begin{figure*}
\centering
\label{fig:Rosette1PROBOBS}
\includegraphics[scale=0.28]{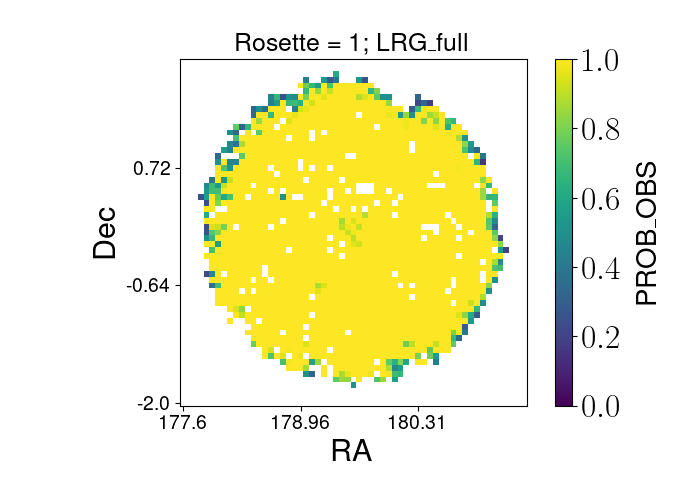}
\includegraphics[scale=0.28]{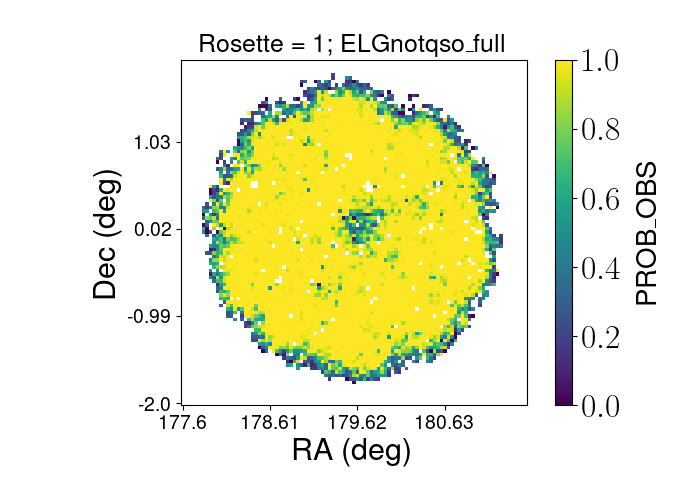}
\includegraphics[scale=0.28]{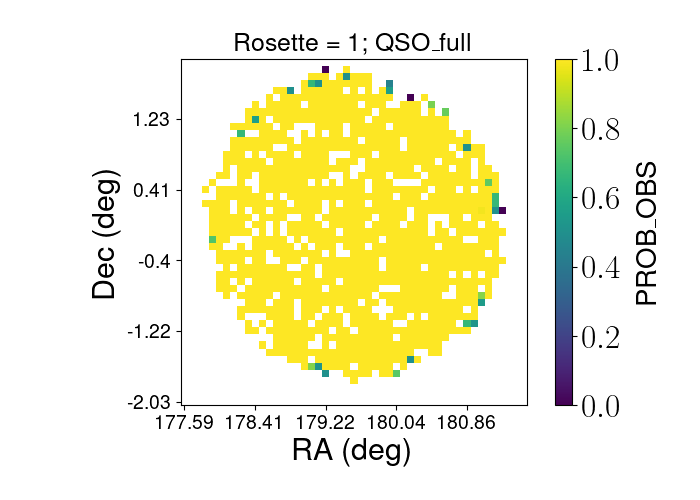}  \caption{The same as Figure \ref{fig:Rosette16PROBOBS} but for Rosette 1} 
\end{figure*}

\begin{figure*}
\centering
\label{fig:Rosette1TargetNumDens}

\includegraphics[scale=0.25]{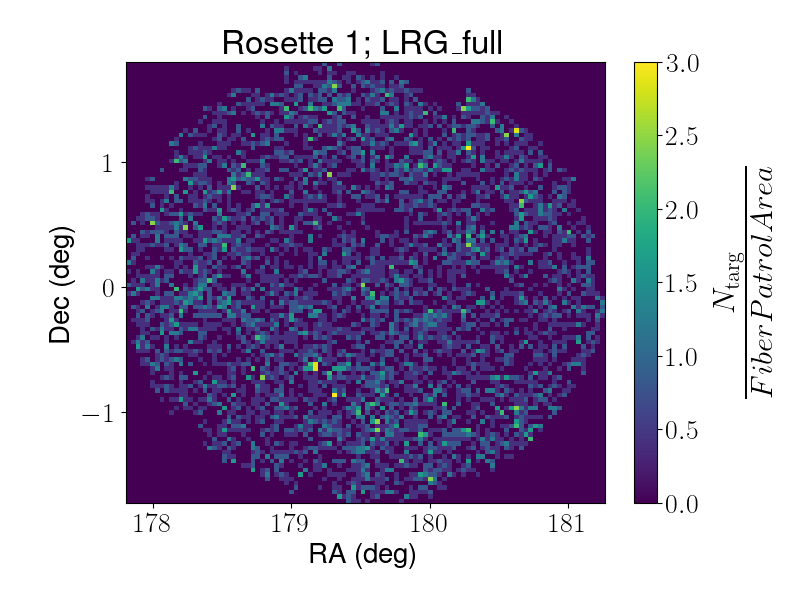}
\includegraphics[scale=0.25]{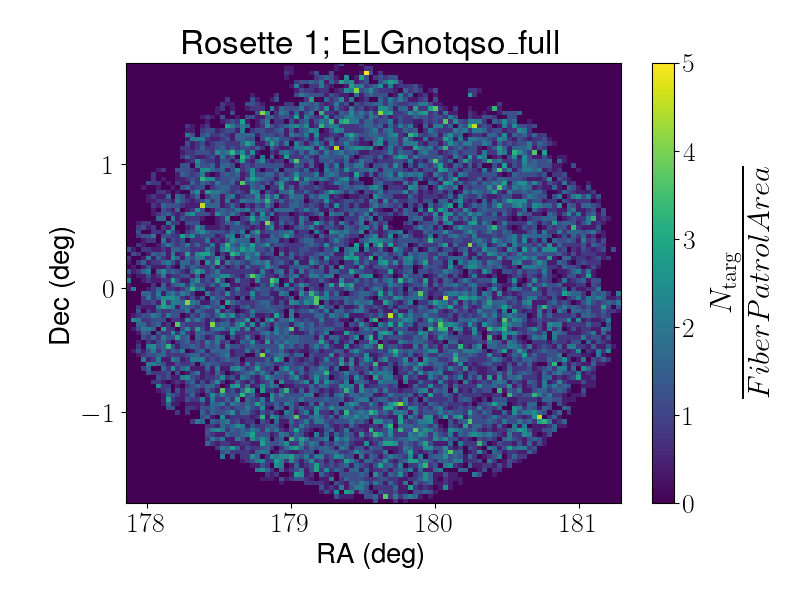}
\includegraphics[scale=0.25]{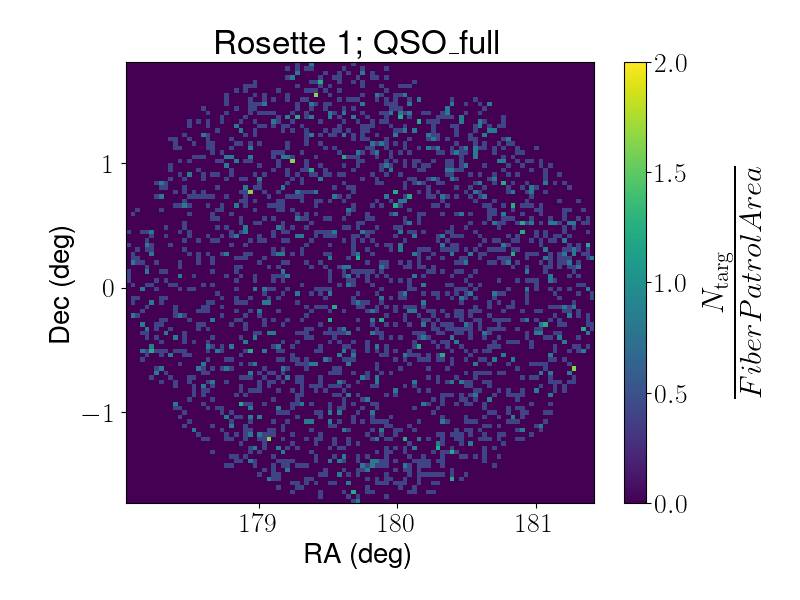}
\caption{The same as Figure \ref{fig:Rosette16TargetNumDens} but for Rosette 1}
\end{figure*}

Figures~\ref{fig:Rosette1PROBOBS} and \ref{fig:Rosette1TargetNumDens} are the equivalents of figures \ref{fig:Rosette16PROBOBS} and \ref{fig:Rosette16TargetNumDens} for rosette 1. We show this rosette as a counterpoint to the rosette 16 since this rosette has much less pronounced differences in areal target number density and represents a more typical rosette. 


\begin{figure*}
\centering
\label{fig:Rosette16Diff}
\includegraphics[scale=0.28]{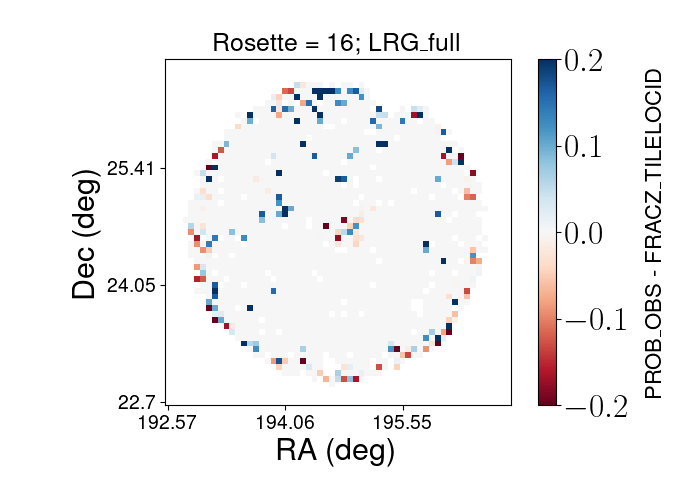}
\includegraphics[scale=0.28]{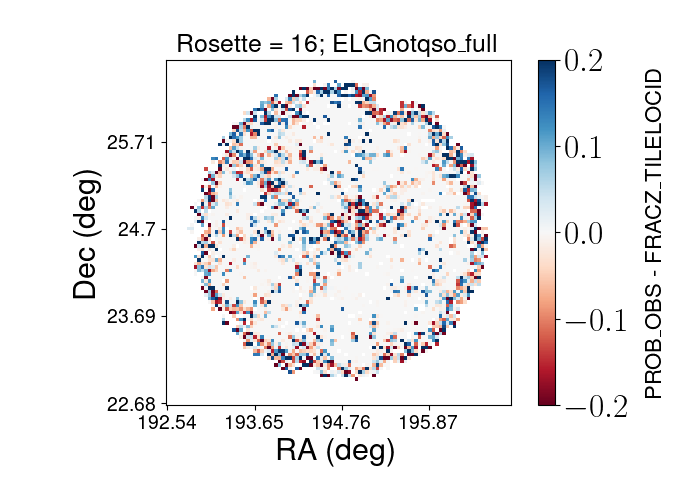}
\includegraphics[scale=0.28]{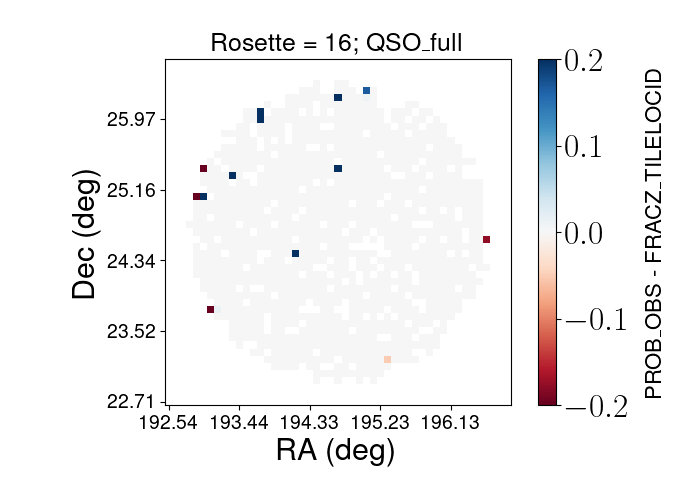}
\caption{2D histogram showing the mean value of the difference between PROB\_OBS and FRACZ\_TILELOCID in bins of RA and Dec within Rosette 16 of SV3 for  (left to right) LRGs, ELGs with QSO targets removed, and QSOs.}
\end{figure*}

\begin{figure*}
\centering
\label{fig:Rosette1Diff}
\includegraphics[scale=0.28]{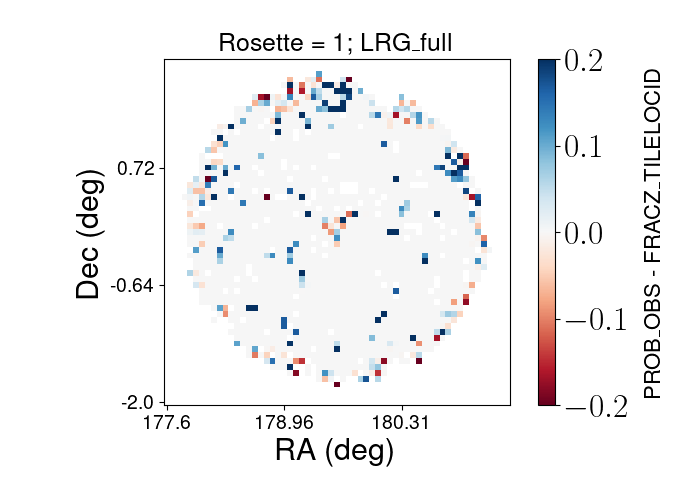}
\includegraphics[scale=0.28]{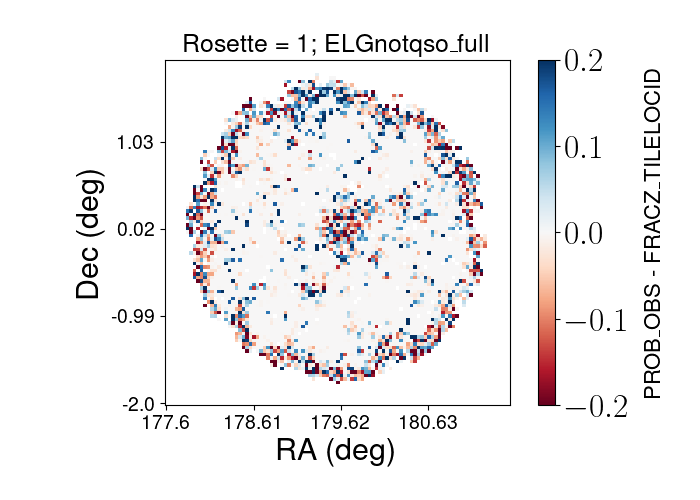}
\includegraphics[scale=0.28]{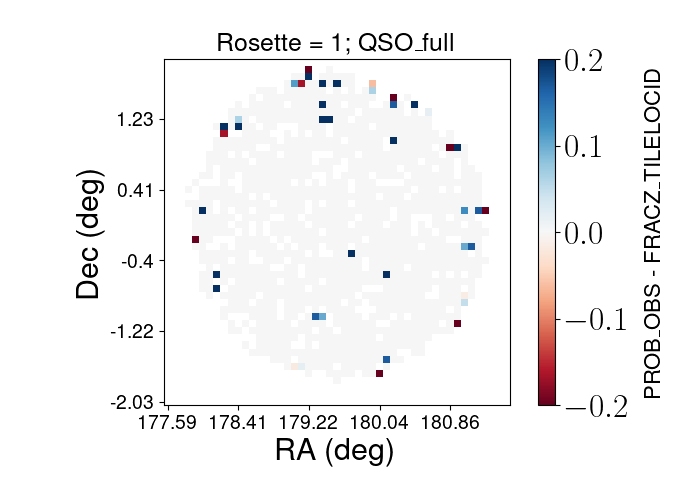}
\caption{The same as Figure \ref{fig:Rosette16Diff} but for Rosette 1.}
\end{figure*}

Figures \ref{fig:Rosette16Diff} and \ref{fig:Rosette1Diff} show the mean value of the difference between PROB\_OBS, the average probability that a target is observed as determined via the \AMTL pipeline presented in this paper and FRACZ\_TILELOCID\footnote{The way SV3 catalogs were constructed \cite{edr}, this is the most appropriate comparison. In the main survey LSS catalog processing, we have defined an additional column \texttt{FRAC\_TLOBS\_TILES}, which accounts for additional incompleteness and should be multiplied by \texttt{FRACZ\_TILELOCID}; see \cite{KP3s15-Ross}. }, the fraction of similar objects within a DESI tile which were successfully observed and received a redshift. These differences show that a utility like the \AMTL pipeline is necessary. A simple monte-carlo simulation of each target independently using what is, at first glance, a reasonable proxy for the probability that it is observed, can result in differences in probability for groups of objects on low completeness pixels of up to, occasionally exceeding, 20\%. 

\subsection{Alternate MTL statistics SV3 mocks}
\begin{table}
\centering
\caption{Number of targets of each class marked as 'observed` in mocks and average sum of PROB\_OBS over 256 mock \AMTL realizations of SV3}
\setlength\tabcolsep{2pt}
\begin{tabular}{cccc}
\label{tab:NOBSTableMock}
Tracer type  & \# 'Observed` & Sum of PROB\_OBS & \%  \\
& in SV3 mocks & in SV3 mocks & difference  \\
\hline
LRG & 140952 & 141995 & 0.73 \\
ELG & 386709  & 388964 & 0.58 \\
QSO & 60129 & 60385 & 0.42 

\end{tabular}
\end{table}
We run the \AMTL pipeline over 25 mocks generated from the AbacusSummit simulations as described in \S \ref{sec:mocks}, reproducing the same algorithm we apply on the observed data using 256 realizations instead of 128. The statistics averaged over the 25 mock realizations, as shown in Table \ref{tab:NOBSTableMock}, show that the number of targets marked as 'observed` in the mocks is $\sim 0.5\%$ smaller than the sum of the PROB\_OBS in the mocks. Due to a larger number of available targets in the mocks, the mocks have significantly more observed targets than the data (shown in Table \ref{tab:NOBSTable}) at a $\sim10\%$ level.

\begin{figure*}
\centering
\label{fig:Rosette16PROBOBSMock}
\includegraphics[scale=0.28]{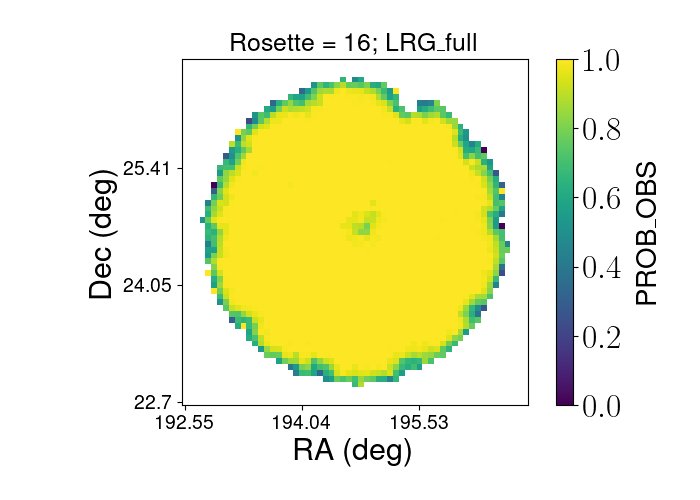}
\includegraphics[scale=0.28]{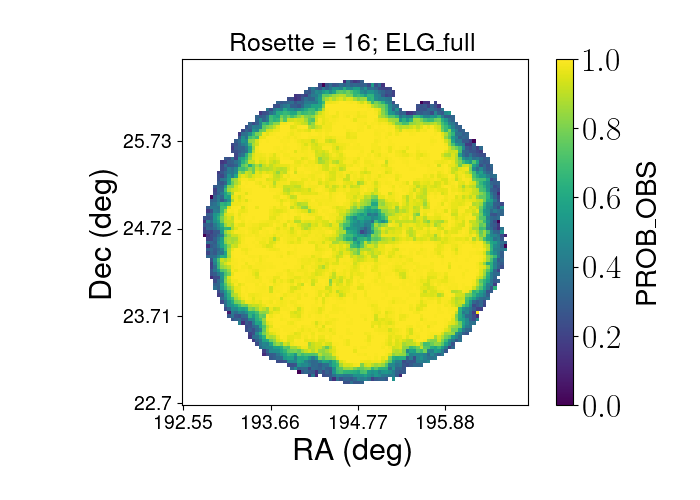}
\includegraphics[scale=0.28]{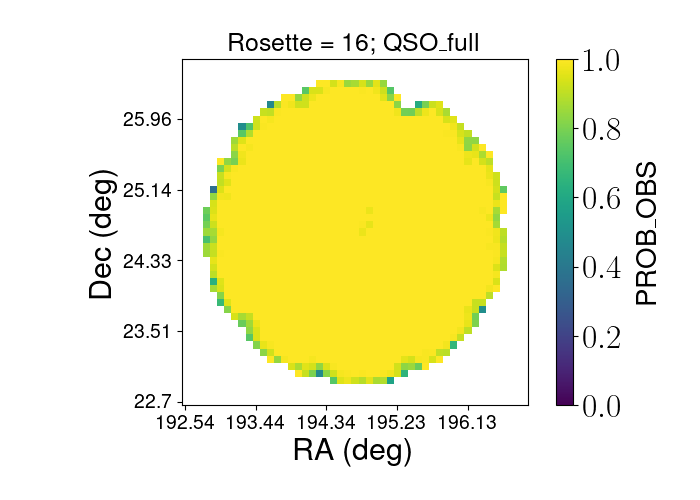}
\caption{2D histogram showing the mean value of PROB\_OBS in bins of RA and Dec from 25 stacked mocks within Rosette 16 of SV3 for (from left to right) LRGs, ELGs, and QSOs.}
\end{figure*}
\begin{figure*}
\centering
\label{fig:Rosette16TargetNumDensMock}
\includegraphics[scale=0.28]{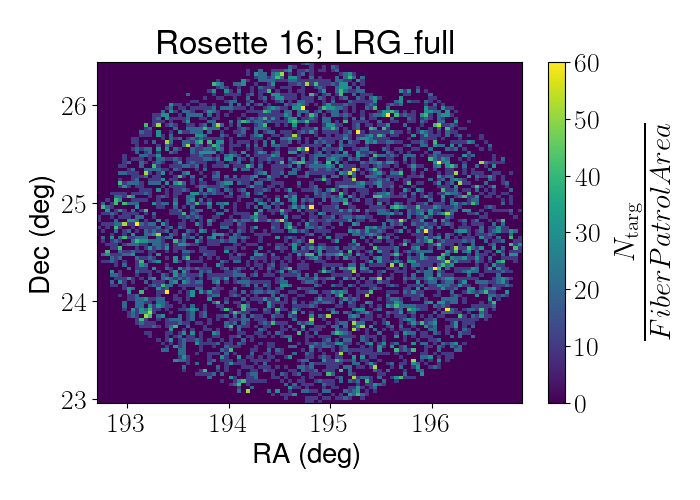}
\includegraphics[scale=0.28]{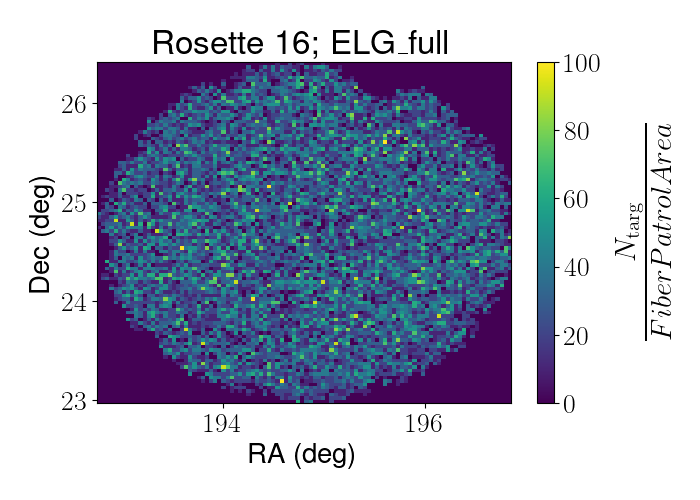}
\includegraphics[scale=0.28]{figures/SV3NumDensPlotNormalizedByFPRLRG_fullRosette16_stackmocks.png}
\caption{2D histogram showing the areal number density (in units of targets per fiber patrol area) of targets in bins of RA and Dec of 25 stacked mocks within Rosette 16 of SV3 for (from left to right) LRGs, ELGs, and  QSOs.}
\end{figure*}
\begin{figure*}
\centering
\label{fig:Rosette1PROBOBSMock}
\includegraphics[scale=0.28]{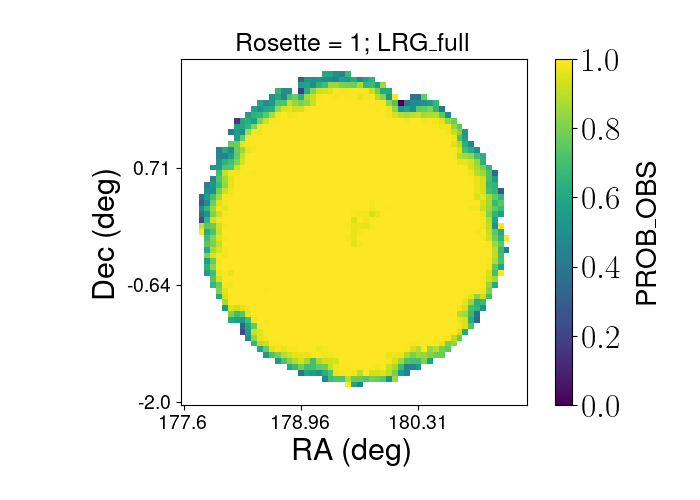}
\includegraphics[scale=0.28]{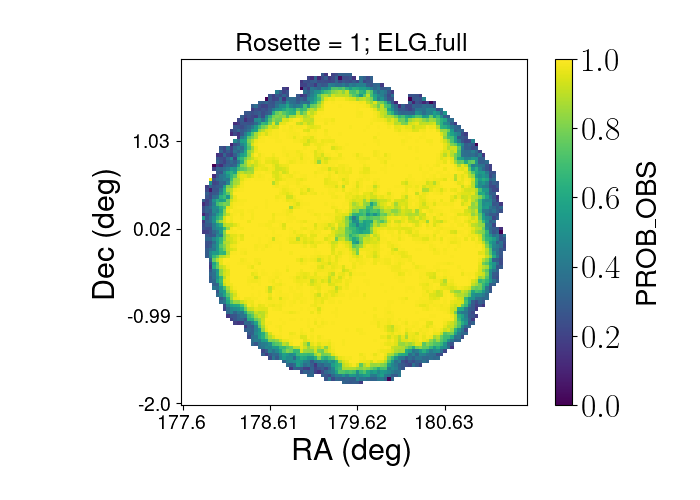}
\includegraphics[scale=0.28]{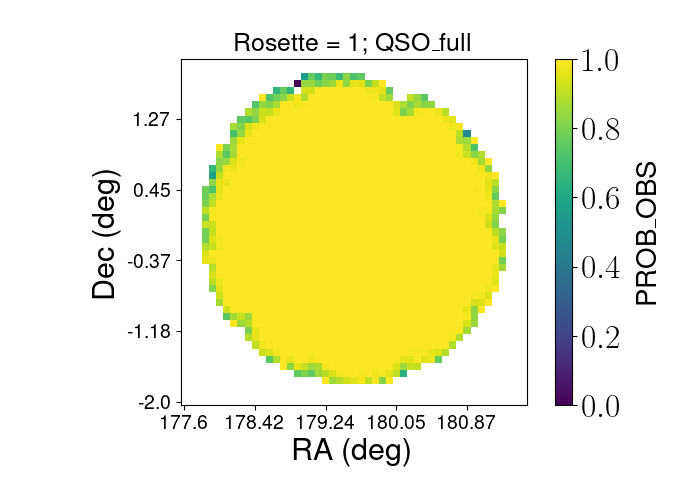}
\caption{The same as Figure~\ref{fig:Rosette16PROBOBSMock}, but for mock rosette 1. }
\end{figure*}
\begin{figure*}
\centering
\label{fig:Rosette1TargetNumDensMock}
\includegraphics[scale=0.28]{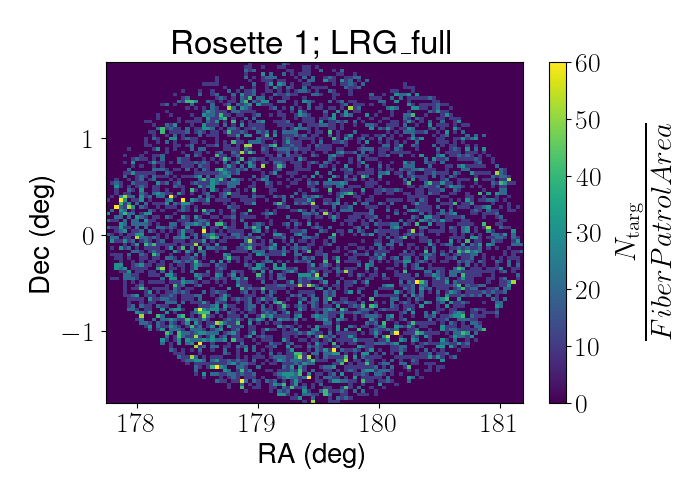}
\includegraphics[scale=0.28]{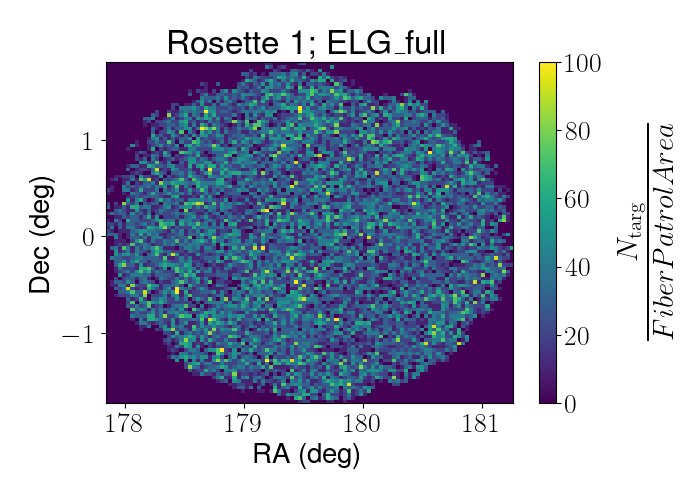}
\includegraphics[scale=0.28]{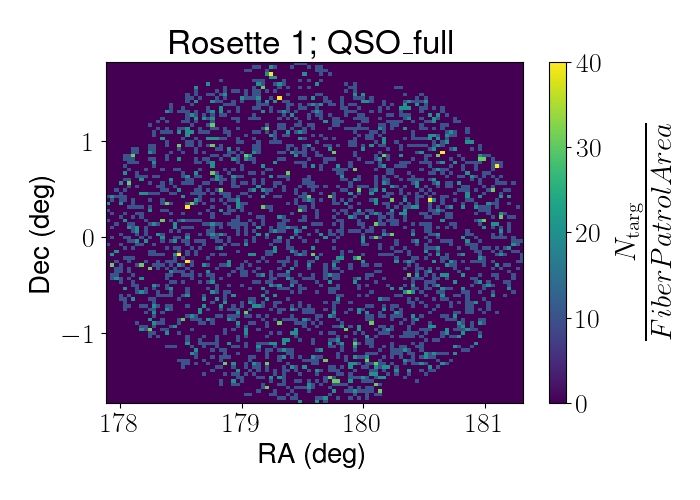}
\caption{The same as Figure~\ref{fig:Rosette16TargetNumDensMock} but for mock Rosette 1.}
\end{figure*}
\begin{figure*}
\centering
\label{fig:Rosette16DiffMock}
\includegraphics[scale=0.28]{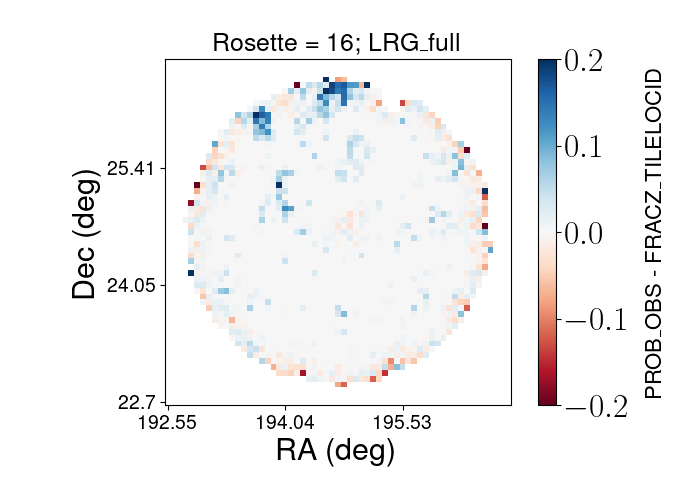}
\includegraphics[scale=0.28]{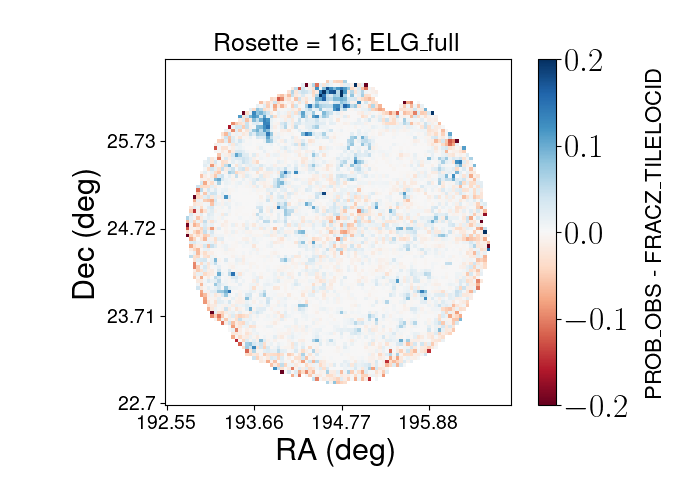}
\includegraphics[scale=0.28]{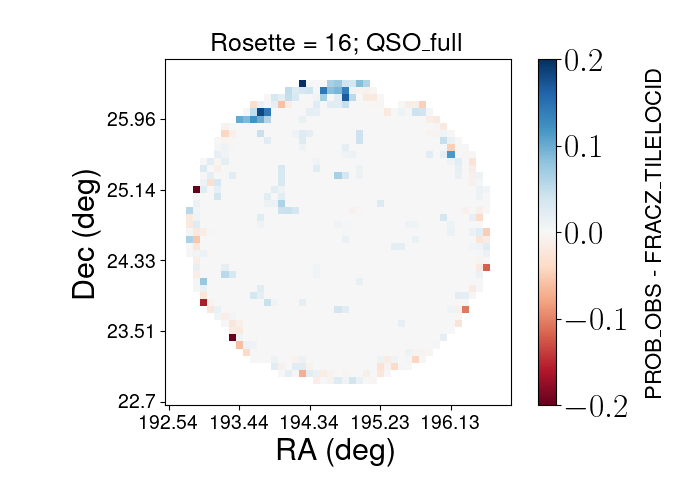}
\caption{2D histogram showing the mean value of the difference between PROB\_OBS and FRACZ\_TILELOCID in bins of RA and Dec from 25 stacked mocks within Rosette 16 of SV3 for (left to right) LRGs,  ELGs , and QSOs.}
\end{figure*}
\begin{figure*}
\centering
\label{fig:Rosette1DiffMock}
\includegraphics[scale=0.28]{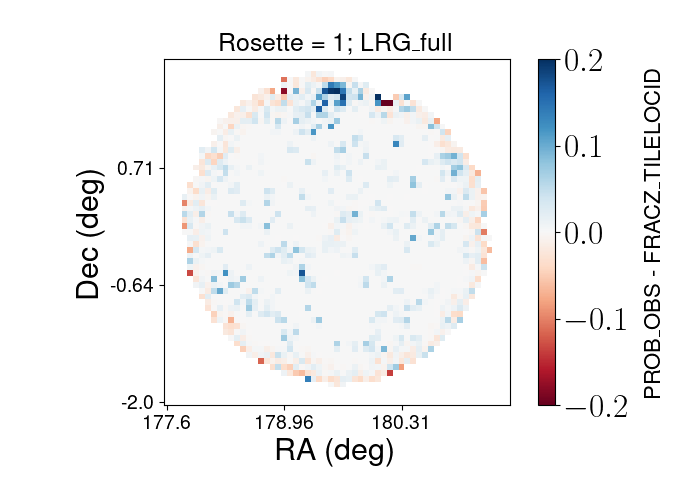}
\includegraphics[scale=0.28]{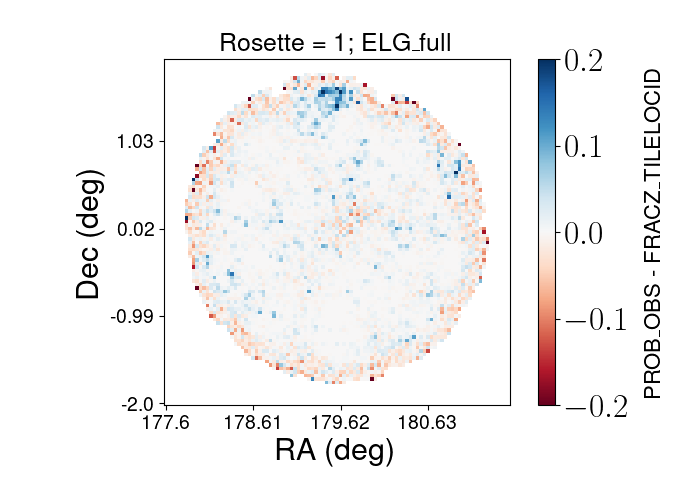}
\includegraphics[scale=0.28]{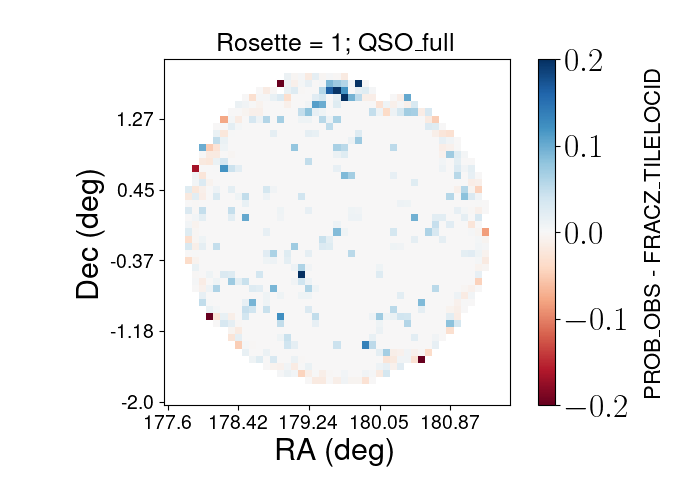}
\caption{The same as Figure~\ref{fig:Rosette16DiffMock} but for Rosette 1}
\end{figure*}

We show the same plots as for the data AMTLs but with a stacked set of all 25 mocks to show that the mock \AMTL process successfully replicates the data PROB\_OBS distribution as well as the difference between PROB\_OBS and FRACZ\_TILELOCID within the limits of the similarity between the mock and data target density.  These plots are the 2D histograms of PROB\_OBS (Figures \ref{fig:Rosette16PROBOBSMock} and \ref{fig:Rosette1PROBOBSMock}), target number density (Figures \ref{fig:Rosette16TargetNumDensMock} and \ref{fig:Rosette1TargetNumDensMock}), and the difference between PROB\_OBS and FRACZ\_TILELOCID (Figures \ref{fig:Rosette16DiffMock} and \ref{fig:Rosette1DiffMock}).


\subsection{SV3 2PCF results}

\begin{figure*}
    \label{fig:MockLRGWPValidation}
    \includegraphics[width=6 in]{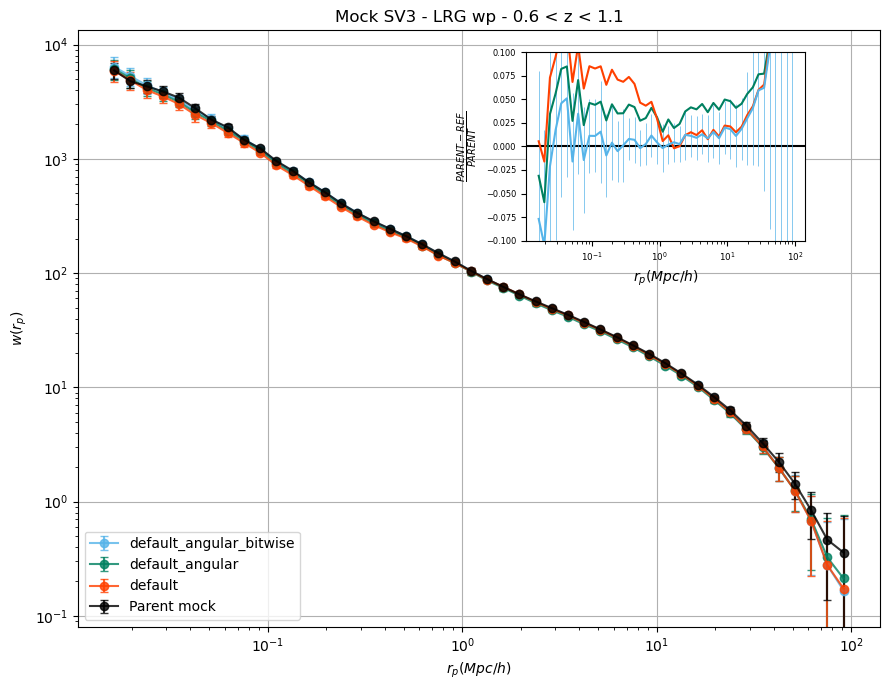}
    \caption{$w_p$ from mean of 25 abacus mocks which were put through the \AMTL pipeline with errorbars from the std. deviation of the 25 $w_p$ measurements. The black line shows the true $w_p$ from the mock parent catalog, the orange line is the observed subset with only default IIP completeness weights, the green line is the observed subset with default IIP completeness weights and the angular upweighting of PB17, and the blue line is the observed subset with angular upweighting and the PIP weights developed in BP17 and computed in this work. The inset shows the fractional difference of the clustering of each of the three 'observed' samples with that of the parent catalog. The PIP weighted measurement has error bars equal to the std. deviation over the 25 mock parent $w_p$ measurements divided by the mean parent $w_p$ measurement. }
\end{figure*}

\begin{figure*}
    \label{fig:MockELGWPValidation}
    \includegraphics[width=6 in]{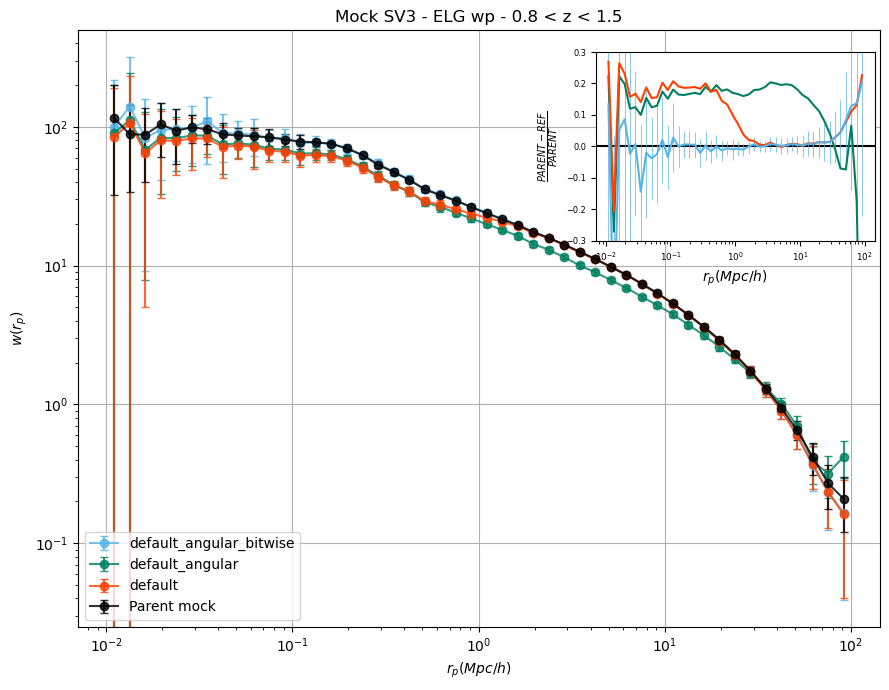}
    \caption{Same as figure \ref{fig:MockLRGWPValidation} for ELGs}
\end{figure*}

\begin{figure*}
    \label{fig:MockQSOWPValidation}
    \includegraphics[width=6 in]{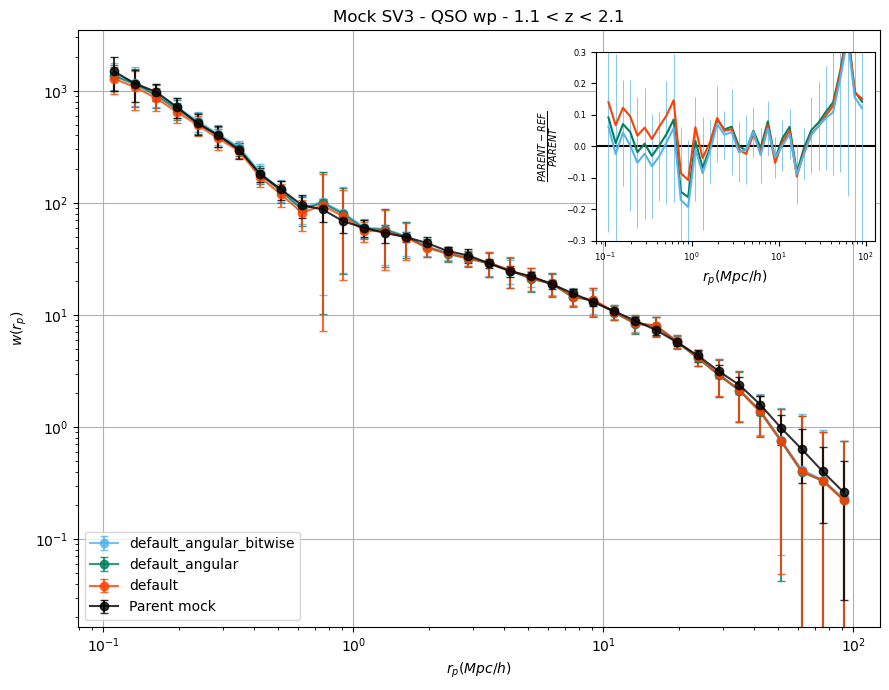}
    \caption{Same as figure \ref{fig:MockLRGWPValidation} for QSOs}
\end{figure*}

Figures \ref{fig:MockLRGWPValidation},  \ref{fig:MockELGWPValidation}, and \ref{fig:MockQSOWPValidation} show the mean projected TPCF over 25 SV3 Abacus mocks with errorbar coming from the standard deviation over the 25 mocks in four cases. The first case, which we treat as truth is the projected TPCF parent catalog of all possible mock targets and is shown in black. The remaining cases are different weights on what was chosen to be the "observed`` set of targets. The "default`` weight (orange) includes the IIP completeness weight, a combination of the default weights and the angular upweighting of PB17 (green; \cite{PercivalAndBianchi} and \S\ref{sec:ClusteringWeights} for more details), and the combination of the angular weights with the bitwise PIP weights discussed here (blue). 

The inset on each plot shows the fractional difference between each weighting scheme with the "truth`` with the blue (PIP weighted) line having error bars equal to the ratio of the "truth`` errorbar and its value in each bin. 

These plots show that the bitweights produced by the \AMTL method are able to recover the parent clustering well within $1-\sigma$ for all tracers down to 0.05 \mpch. Furthermore, the PIP weighting improves the recovery of the parent catalog clustering in all but a few separation bins. 

The case of ELGs is most informative, since ELGs are the least complete target class. Clustering measured using only the default weighting is off by 20\% uniformly down to a scale of nearly 1\mpch. Additionally, the addition of angular upweights causes a spurious shift which pushes the 20\% discrepancy down to a scale over 10 \mpch. The PIP weights when compbined with the above schemes shows agreement within a few percent between scales of 0.05 \mpch and 25 \mpch with agreement within $1-\sigma$ at all scales shown.

\section{Discussion}
\label{sec:discussion}

Comparing the statistics of the data \AMTL with the data shows a small, but persistent $\sim 1\%$ excess in the number of observed targets. The mocks also show a similar excess. This is likely due to the \AMTL pipeline not fully accurately sampling observation results from the data for the alternate targets. As mentioned in section \S\ref{sec:ObsLoop}, we take observation results in the \AMTL from the exact same fiber in the observations without regard to the target type which is assigned in the alternate realizations. In a future release of the \AMTL pipeline we will have a fiber twin method implemented which will select observation results from a nearby fiber with nearly identical observational properties including target type.  
While the number of total observations shows some disagreements, the shape of the PROB\_OBS distribution is recovered very well. For Rosette 16, the data PROB\_OBS distributions in Figure~\ref{fig:Rosette16PROBOBS} and mock PROB\_OBS distributions in Figure~\ref{fig:Rosette16PROBOBSMock} show nearly identical structure with the mocks showing much less fluctuation due to higher statistics and the data showing a small decrease in PROB\_OBS towards the lower right of the rosette not present in the mocks due to a high target number density in that region. Rosette 1 also shows good agreement between Figures~\ref{fig:Rosette1PROBOBS} and \ref{fig:Rosette1PROBOBSMock} with the significantly higher PROB\_OBS in the top left of the data rosette as compared to the mock coming from the very low data target number density. 

The difference between PROB\_OBS and the more generic FRACZ\_TILELOCID in figures~\ref{fig:Rosette16Diff}, \ref{fig:Rosette1Diff}, \ref{fig:Rosette16DiffMock}, and \ref{fig:Rosette1DiffMock} show similar patterns. There is very little difference between the two completeness metrics in the centers of both rosettes since completeness is at or near 100\% at all locations. Towards the edges, we see a significant amount of fluctuation between the PROB\_OBS and FRACZ\_TILELOCID suggesting that completeness is somewhat different when measured through more approximate methods than an \AMTL method. 

Figures~\ref{fig:MockLRGWPValidation}, \ref{fig:MockELGWPValidation}, and \ref{fig:MockQSOWPValidation} show the validation of the \AMTL method by using the bitweights obtained by using the method on mocks to recover the clustering of the parent mocks. Since the true redshift of all targets in the mock are known, this enables us to validate against a known "truth`` projected two point correlation function. 

The validation results show that the bitwise weights calculated from the \AMTL method are a strict improvement over the default completeness weights at small scales. For all three dark time tracers, the (blue) PIP+ANG upweighted mock projected two point correlation functions are an improvement of the default completeness weights at nearly all small ( $\leq 1 $\mpch ) scales.  Furthermore, down to scales of 0.02, 0.01, and 0.1 \mpch (for LRGs, ELGs, and QSOs respectively), the PIP+ANG upweighted clustering from the mocks is consistent with the parent mock clustering within the scatter among the 25 mocks. These results, especially the ELG clustering, show that the PIP weighting and angular upweighting are both necessary to correct the clustering measurements for incompleteness. For a sample like SV3 which is highly complete, IIP weights function similarly to nearest neighbor weights, indicating that those weights would be insufficient for incompleteness correction at small scales. 

The larger scale clustering ($s \geq 1 $ \mpch and especially $s \geq 10 $\mpch), while statistically consistent with the parent clustering within the mock variance, is not as well recovered. However, there are several reasons why this should be the case. First, the SV3 rosette design strongly limits the number of pairs which we can observe at separations nearing 100 \mpch. The size of an individual rosette at a redshift z=1.0 is only $\sim 70 $\mpch. While DESI  did place some rosettes adjacent to each other enabling the observation of some pairs at those separations, the vast majority of rosettes were isolated.

In conclusion, we have demonstrated the generation of 128 realizations of DESI SV3 and Y1 of the main survey using a nearly-maximally-realistic method utilizing alternative merged target ledgers. We have validated the \AMTL method on 256 realizations of DESI SV3-like mocks showing the recovery of clustering down to scales of 0.01 \mpch for the least complete tracer, Emission Line Galaxies.

\section*{Acknowledgements}

JL and RK acknowledge support of the U.S. Department of Energy (DOE) in funding grant DE-SC0010129 for the work in this paper. Computational resources for JL were provided by SMU’s Center for Research Computing. This work made use of Astropy:\footnote{http://www.astropy.org} a community-developed core Python package and an ecosystem of tools and resources for astronomy \citep{astropy:2013, astropy:2018, astropy:2022}
ADM was supported by the U.S.\ Department of Energy, Office of Science, Office of High Energy Physics, under Award Number DE-SC0019022.

This material is based upon work supported by the U.S. Department of Energy (DOE), Office of Science, Office of High-Energy Physics, under Contract No. DE–AC02–05CH11231, and by the National Energy Research Scientific Computing Center, a DOE Office of Science User Facility under the same contract. Additional support for DESI was provided by the U.S. National Science Foundation (NSF), Division of Astronomical Sciences under Contract No. AST-0950945 to the NSF’s National Optical-Infrared Astronomy Research Laboratory; the Science and Technology Facilities Council of the United Kingdom; the Gordon and Betty Moore Foundation; the Heising-Simons Foundation; the French Alternative Energies and Atomic Energy Commission (CEA); the National Council of Humanities, Science and Technology of Mexico (CONAHCYT); the Ministry of Science and Innovation of Spain (MICINN), and by the DESI Member Institutions: \url{https://www.desi.lbl.gov/collaborating-institutions}. Any opinions, findings, and conclusions or recommendations expressed in this material are those of the author(s) and do not necessarily reflect the views of the U. S. National Science Foundation, the U. S. Department of Energy, or any of the listed funding agencies.

The authors are honored to be permitted to conduct scientific research on Iolkam Du’ag (Kitt Peak), a mountain with particular significance to the Tohono O’odham Nation.

\section{Data Availability}
The SV3 data used in this analysis is currently availabile to the public at https://data.desi.lbl.gov/doc/releases/edr/ .
The Y1 data used in this analysis will be made public along the Data Release 1 (details in https://data.desi.lbl.gov/doc/releases/)
\bibliographystyle{unsrtnat}
\bibliography{AltMTL}

\appendix
\section{Author Affiliations}
\label{sec:affiliations}

\begin{hangparas}{.5cm}{1}

$^{1}${Department of Physics, Southern Methodist University, 3215 Daniel Avenue, Dallas, TX 75275, USA}

$^{2}${Departamento de Astrof\'{\i}sica, Universidad de La Laguna (ULL), E-38206, La Laguna, Tenerife, Spain}

$^{3}${Instituto de Astrof\'{\i}sica de Canarias, C/ V\'{\i}a L\'{a}ctea, s/n, E-38205 La Laguna, Tenerife, Spain}

$^{4}${Department of Physics \& Astronomy, University  of Wyoming, 1000 E. University, Dept.~3905, Laramie, WY 82071, USA}

$^{5}${Center for Cosmology and AstroParticle Physics, The Ohio State University, 191 West Woodruff Avenue, Columbus, OH 43210, USA}

$^{6}${Department of Astronomy, The Ohio State University, 4055 McPherson Laboratory, 140 W 18th Avenue, Columbus, OH 43210, USA}

$^{7}${The Ohio State University, Columbus, 43210 OH, USA}

$^{8}${Dipartimento di Fisica ``Aldo Pontremoli'', Universit\`a degli Studi di Milano, Via Celoria 16, I-20133 Milano, Italy}

$^{9}${University of Michigan, Ann Arbor, MI 48109, USA}

$^{10}${IRFU, CEA, Universit\'{e} Paris-Saclay, F-91191 Gif-sur-Yvette, France}

$^{11}${Department of Physics and Astronomy, University of Waterloo, 200 University Ave W, Waterloo, ON N2L 3G1, Canada}

$^{12}${Perimeter Institute for Theoretical Physics, 31 Caroline St. North, Waterloo, ON N2L 2Y5, Canada}

$^{13}${Waterloo Centre for Astrophysics, University of Waterloo, 200 University Ave W, Waterloo, ON N2L 3G1, Canada}

$^{14}${Lawrence Berkeley National Laboratory, 1 Cyclotron Road, Berkeley, CA 94720, USA}

$^{15}${Physics Dept., Boston University, 590 Commonwealth Avenue, Boston, MA 02215, USA}

$^{16}${Institute for Computational Cosmology, Department of Physics, Durham University, South Road, Durham DH1 3LE, UK}

$^{17}${Department of Physics \& Astronomy, University College London, Gower Street, London, WC1E 6BT, UK}

$^{18}${Instituto de F\'{\i}sica, Universidad Nacional Aut\'{o}noma de M\'{e}xico,  Cd. de M\'{e}xico  C.P. 04510,  M\'{e}xico}

$^{19}${Department of Astronomy, School of Physics and Astronomy, Shanghai Jiao Tong University, Shanghai 200240, China}

$^{20}${Kavli Institute for Particle Astrophysics and Cosmology, Stanford University, Menlo Park, CA 94305, USA}

$^{21}${SLAC National Accelerator Laboratory, Menlo Park, CA 94305, USA}

$^{22}${Departamento de F\'isica, Universidad de los Andes, Cra. 1 No. 18A-10, Edificio Ip, CP 111711, Bogot\'a, Colombia}

$^{23}${Observatorio Astron\'omico, Universidad de los Andes, Cra. 1 No. 18A-10, Edificio H, CP 111711 Bogot\'a, Colombia}

$^{24}${Institut d'Estudis Espacials de Catalunya (IEEC), 08034 Barcelona, Spain}

$^{25}${Institute of Cosmology and Gravitation, University of Portsmouth, Dennis Sciama Building, Portsmouth, PO1 3FX, UK}

$^{26}${Institute of Space Sciences, ICE-CSIC, Campus UAB, Carrer de Can Magrans s/n, 08913 Bellaterra, Barcelona, Spain}

$^{27}${Fermi National Accelerator Laboratory, PO Box 500, Batavia, IL 60510, USA}

$^{28}${Department of Physics, The Ohio State University, 191 West Woodruff Avenue, Columbus, OH 43210, USA}

$^{29}${School of Mathematics and Physics, University of Queensland, 4072, Australia}

$^{30}${NSF NOIRLab, 950 N. Cherry Ave., Tucson, AZ 85719, USA}

$^{31}${Sorbonne Universit\'{e}, CNRS/IN2P3, Laboratoire de Physique Nucl\'{e}aire et de Hautes Energies (LPNHE), FR-75005 Paris, France}

$^{32}${Departament de F\'{i}sica, Serra H\'{u}nter, Universitat Aut\`{o}noma de Barcelona, 08193 Bellaterra (Barcelona), Spain}

$^{33}${Institut de F\'{i}sica d’Altes Energies (IFAE), The Barcelona Institute of Science and Technology, Campus UAB, 08193 Bellaterra Barcelona, Spain}

$^{34}${Instituci\'{o} Catalana de Recerca i Estudis Avan\c{c}ats, Passeig de Llu\'{\i}s Companys, 23, 08010 Barcelona, Spain}

$^{35}${Department of Physics and Astronomy, Siena College, 515 Loudon Road, Loudonville, NY 12211, USA}

$^{36}${Department of Physics and Astronomy, University of Sussex, Brighton BN1 9QH, U.K}

$^{37}${National Astronomical Observatories, Chinese Academy of Sciences, A20 Datun Rd., Chaoyang District, Beijing, 100012, P.R. China}

$^{38}${Departamento de F\'{i}sica, Universidad de Guanajuato - DCI, C.P. 37150, Leon, Guanajuato, M\'{e}xico}

$^{39}${Instituto Avanzado de Cosmolog\'{\i}a A.~C., San Marcos 11 - Atenas 202. Magdalena Contreras, 10720. Ciudad de M\'{e}xico, M\'{e}xico}

$^{40}${Korea Astronomy and Space Science Institute, 776, Daedeokdae-ro, Yuseong-gu, Daejeon 34055, Republic of Korea}

$^{41}${Space Sciences Laboratory, University of California, Berkeley, 7 Gauss Way, Berkeley, CA  94720, USA}

$^{42}${University of California, Berkeley, 110 Sproul Hall \#5800 Berkeley, CA 94720, USA}

$^{43}${Instituto de Astrof\'{i}sica de Andaluc\'{i}a (CSIC), Glorieta de la Astronom\'{i}a, s/n, E-18008 Granada, Spain}

$^{44}${Department of Physics, Kansas State University, 116 Cardwell Hall, Manhattan, KS 66506, USA}

$^{45}${Department of Physics and Astronomy, Sejong University, Seoul, 143-747, Korea}

$^{46}${CIEMAT, Avenida Complutense 40, E-28040 Madrid, Spain}

$^{47}${Department of Physics, University of Michigan, Ann Arbor, MI 48109, USA}

$^{48}${Department of Physics \& Astronomy, Ohio University, Athens, OH 45701, USA}

\end{hangparas}

\end{document}